\documentclass[prc,showpacs,showkeys,nofootinbib,superscriptaddress]{revtex4}
\usepackage{subfig}
\usepackage{graphicx}
\usepackage{dcolumn}
\usepackage{epsfig}
\usepackage{bm}
\newcommand{\eq}{\begin{eqnarray}}
\newcommand{\en}{\end{eqnarray}}

\newcommand{\ba}[1]{\begin{eqnarray} \label{(#1)}}
\newcommand{\ea}{\end{eqnarray}}

\newcommand{\newc}{\newcommand}
\def\vbf{\mbox{\boldmath $\upsilon$}}
\newc{\lra}{\leftrightarrow}
\newc{\beq}{\begin{equation}}
\newc{\eeq}{\end{equation}}
\newc{\barr}{\begin{eqnarray}}
\newc{\earr}{\end{eqnarray}}
  
\begin{document}

\topmargin -0.50in
\title {Debris Flows in Direct Dark Matter Searches:The modulation effect}

\author{J.D. Vergados}
\affiliation{Theoretical Physics Division, University of Ioannina, 
GR--451 10 Ioannina, Greece.}
\begin{abstract}
The effect of some possible non standard WIMP velocity distributions, like the Debris Flows recently proposed, on the direct dark matter detection rates is investigated. We find that such distributions may be deciphered from the data, especially if the time variation of the event rates due to the annual motion of the Earth is observed.
\end{abstract}



\date{\today}

\maketitle
\section{Introduction}
Combining the data of all the available observations and, in particular, the data of the precise experiments \cite{SPERGEL,WMAP06}, we now know that most of the matter in the Universe is dark, i.e. exotic and non baryonic.
Furthermore there exists firm indirect evidence for a halo of dark matter
in galaxies from the
observed rotational curves, see e.g the review \cite{UK01}.  It is, however, essential to directly
detect any
such dark matter,  a task, which, of course, depends on the nature of the dark matter
constituents or WIMPs (weakly interacting massive particles) and their interactions.

Since the WIMPs are  expected to be
extremely non relativistic, with average kinetic energy $\langle T\rangle  \approx
50 \mbox{ keV } (m_{\mbox{{\tiny WIMP}}}/ 100 \ {\rm GeV} )$, they are not likely to excite the nucleus.
So, they can be directly detected mainly via the recoiling of the target nucleus
(A,Z) following the WIMP-nucleus scattering. The event rate for such a process can
be computed from the following ingredients \cite{LS96}:
i) The elementary WIMP-nucleon scattering cross section. This most important parameter will not, however, be the subject of the present work. We will adopt the view that it   can be extracted from the data of event rates, if and when such data become available.
ii) Knowledge of the relevant nuclear matrix elements
 obtained with as reliable as possible many
body nuclear wave functions. In the present work we will limit ourselves to elastic WIMP-nucleus scattering and, thus, only the nuclear form factor is needed. iii) Knowledge of the WIMP density in our vicinity and its velocity distribution.

In the standard nuclear recoil experiments, first proposed more than 30 years ago \cite{GOODWIT}, one has to face the
 problem that the reaction of interest does not have a characteristic feature to distinguish it
from the background. So for the expected low counting rates the background is
a formidable problem. Some special features of the WIMP-nuclear interaction can be exploited to 
reduce the background problems. Such as:

i) The modulation effect. This yields a periodic signal due to the motion of the Earth around the Sun. This effect, also proposed a long time ago \cite{Druck} and subsequently studied by many authors \cite{PSS88,GS93,RBERNABEI95,LS96,ABRIOLA98,HASENBALG98,JDV03,GREEN04,SFG06,FKLW11}, depends on the assumed velocity distribution. In the standard Maxwell Boltzmann (M-B) distribution for WIMPs in the Galactic halo the  modulation amplitude, 
depending on the mechanism of the reaction as well as on the target and the WIMP mass, is small. The relative amplitude  becomes even smaller than  $2\%$ in the case of low detector energy cut off \cite{JDV03}. 

ii) Backward-forward asymmetry expected in directional experiments, i.e. experiments in which the direction of the recoiling nucleus is also observed. Such an asymmetry has also been predicted a long time ago \cite{SPERGEL88}, but it has not yet been exploited, since such experiments have been considered  very difficult to perform.
Some progress has, however, recently been made in this direction and   the relevant experiments  now appear  feasible \cite{SPERGEL88,DRIFT,SHIMIZU03,KUDRY04,DRIFT2,GREEN05,Green06,KRAUSS,KRAUSS01,Alenazi08,Creswick010,Lisanti09,Giometal11}. In such experiments the event rate and its modulation depend on the direction of observation.

 An essential ingredient in direct WIMP detection is the WIMP density in our vicinity and, especially, the WIMP velocity distribution. The dark matter in the solar neighborhood is commonly assumed to be smoothly distributed in
space and to have a Maxwellian velocity distribution. Some of the calculations have considered various forms of phenomenological non symmetric velocity distributions  \cite{DRIFT2,GREEN04,GREEN05,VEROW06,JDV09,TETRVER06,VSH08}  and some of them even more exotic dark matter flows like
the late infall of dark matter into  the Galaxy, i.e caustic rings
 \cite{SIKIVI1,SIKIVI2,Verg01,Green,Gelmini} and Sagittarius dark matter \cite{GREEN02}.
 
In addition to the above models very recently it was found that the velocity distributions measured in high resolution numerical simulations exhibit deviations from the standard Maxwell-Boltzmann assumption, especially at large velocities \cite{KUHLEN10,LAWW11}. Furthermore a distinction was  made between a velocity structure that is spatially localized, such as 
streams \cite{SBWMZ08,PKB09}, and that which is spatially homogenized, which was  designated as ``debris flow'' \cite{LisSper11}. Both
streams  and debris flows
 arise from the disruption of satellites that fall into the Milky Way, but differ in the relative amount of phase-mixing that they have undergone. Implications of streams \cite{streams11} and, more recently, of  the debris flows in direct dark matter searches have  been considered by  Kuhlen, Lisanti and Spergel \cite{spergel12}.
 
 In the present paper we will discuss in some detail the effect of these debris flows\cite{spergel12} on the event rates of direct dark matter experiments  for a variety of targets  such as those employed in XENON10 \cite{XENON10}, XENON100 \cite{XENON100.11}, XMASS \cite{XMASS09}, ZEPLIN \cite{ZEPLIN11}, PANDA-X \cite{PANDAX11}, LUX \cite{LUX11}, CDMS \cite{CDMS05}, CoGENT \cite{CoGeNT11}, EDELWEISS \cite{EDELWEISS11}, DAMA \cite{DAMA1,DAMA11}, KIMS \cite{KIMS07} and PICASSO \cite{PICASSO09,PICASSO11}. We will also study the effect of these flows on the time variation of the relevant rates due to the annual motion of the Earth \cite{JDV12n} (modulation effect) as a function of the energy transfer and the WIMP mass and compare them with the standard M-B distribution. The effects of debris flows in directional experiments will be studied elsewhere.


\section{The formalism for the WIMP-nucleus differential event rate}
Before calculating the direct detection event rate, we will first deal with the WIMP velocity distribution. To this end we will follow the steps:
\begin{itemize}
\item One starts with such distribution in the Galactic frame.
\item one transforms to the local coordinate system:
\beq
{\bf y} \rightarrow {\bf y}+{\hat\upsilon}_s+\delta \left (\sin{\alpha}{\hat x}-\cos{\alpha}\cos{\gamma}{\hat y}+\cos{\alpha}\sin{\gamma} {\hat \upsilon}_s\right ) ,\quad y=\frac{\upsilon}{\upsilon_0}
\label{Eq:vlocal}
\eeq
with $\gamma\approx \pi/6$, $ {\hat \upsilon}_s$ a unit vector in the Sun's direction of motion, $\hat{x}$  a unit vector radially out of the galaxy in our position and  $\hat{y}={\hat \upsilon}_s\times \hat{x}$. The last term in the first expression of Eq. (\ref{Eq:vlocal}) corresponds to the motion of the Earth around the Sun with $\delta$ being the ratio of modulus of the Earth's velocity around the Sun divided by the Sun's velocity around the center of the Galaxy, i.e.  $\upsilon_0\approx 220$km/s and $\delta\approx0.135$. The above formula assumes that the motion  of both the Sun around the Galaxy and of the Earth around the Sun are uniformly circular. The exact orbits are, of course, more complicated \cite{GREEN04,LANG99}, but such deviations are not expected to significantly modify our results. In Eq. (\ref{Eq:vlocal}) $\alpha$ is the phase of the Earth ($\alpha=0$ around June 3nd)\footnote{One could, of course, make the time dependence of the rates due to the motion of the Earth more explicit by writing $\alpha \approx(6/5)\pi\left (2 (t/T)-1 \right )$, where $t/T$ is the fraction of the year.}.
\item One integrates  the velocity distribution over the angles and the result is multiplied by the  velocity $\upsilon$ due to the WIMP flux.
\item The result is integrated from a minimum value $\upsilon_{min}$ to the maximum allowed velocity  $\upsilon_{max}$. In general, the escape velocity $ \upsilon_{esc}$ in our galaxy is estimated to be in the range  550km/s$\le\upsilon_{esc}\le650$km/s. In our calculations we assumed for the M-B distribution  $\upsilon_{esc}\approx620$km/s,  even though the value\cite{spergel12} of 550 km/s, which results from an analysis of data from the RAVE survey \cite{RAVE06}, would have been more appropriate. The obtained results are not sensitive to this value. $\upsilon_{min}$ is a suitable parametrization in terms of the recoil energy and the target parameters, namely:
\beq
\upsilon_{min}=\sqrt{\frac{A m_p E_R}{2 \mu^2_r}}
\eeq
 where $A m_p$ is the mass of the nucleus, $\mu_r $ is the reduced mass of the WIMP-nucleus system and $E_R$ is the energy transfer to the nucleus. 
\end{itemize}

\begin{figure}
\begin{center}
\rotatebox{90}{\hspace{0.0cm} $g(\upsilon_{min},\upsilon_E(\alpha))\times10^{3}\rightarrow$(km/s)$^{-1}$}
\includegraphics[height=0.4\textwidth]{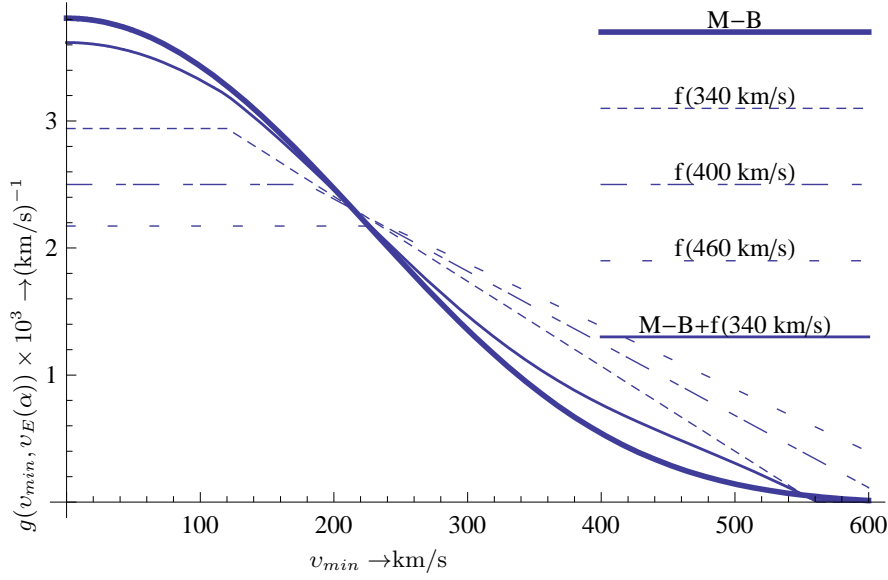}
\\
{\hspace{-2.0cm} $\upsilon_{min}\rightarrow$km/s}
\end{center}
\caption{ The  function $g(\upsilon_{min})$ as a function of $\upsilon_{min}$ in the local frame considered in this work in the case of  the traditional M-B distribution as well as the debris flows of Kuhlen, Lisanti and Spergel \cite{spergel12}. The flows presented here are  indicated by the symbol f followed in parenthesis by the  flow velocity in the Galactic frame. The combination of M-B and the indicated  flow was obtained via Eqs (\ref{Eq:epsilonmix}) and (\ref{Eq:M-BFlow}) below.
 \label{fig:flowv}}
\end{figure}

With the above procedure  one obtains the quantity  $g(\upsilon_{min})$. For the M-B distribution in the local frame it is defined as follows:
\beq
g(\upsilon_{min},\upsilon_E(\alpha))=\frac{1}{\left (\sqrt{\pi}\upsilon_0 \right )^3}\int_{\upsilon_{min}}^{\upsilon_{max}}e^{-(\upsilon^2+2 \vbf . \vbf_E(\alpha)+\upsilon_E^2(\alpha))/\upsilon^2_0}\upsilon d\upsilon d \Omega,\quad\upsilon_{max}=\upsilon_{esc}
\eeq 
$\vbf_E(\alpha)$ is the velocity of the Earth around the Sun (see Eq. (\ref{Eq:vlocal})). The above upper cut off value in the M-B is usually put in by hand. Such a cut off comes in naturally, however, in the case of velocity distributions obtained from the halo WIMP mass density in the Eddington approach \cite{VEROW06}, which, in certain models, resemble a M-B distribution \cite{JDV09}. Its precise value is not, however, important for the results of the present paper (see Fig. \ref{Fig:effyesc} below).\\
For the isotropic debris flows considered by  Kuhlen, Lisanti and Spergel \cite{spergel12} it is given by:
\beq
g(\upsilon_{min},\upsilon_E(\alpha))=\int_{\upsilon_{min}}\frac{f(\upsilon)}{\upsilon}d \upsilon, \, f(\upsilon)= \left \{\begin{array}{ll}\frac{\upsilon}{2
 \upsilon_{flow}\upsilon_E(\alpha)},&\upsilon_{flow}-\upsilon_E(\alpha)<\upsilon<\upsilon_{flow}+\upsilon_E(\alpha)\\
0,&\mbox{otherwise} 
\end{array}\right .
\eeq
where $\upsilon_{flow}$ is the flow velocity in the Galactic frame. These functions are  shown in Fig. \ref{fig:flowv} for the M-B distribution and the debris flows indicated by the symbol f followed by the flow velocity in the Galactic frame enclosed in parenthesis. A combination of M-B and a particular flow, obtained via Eqs (\ref{Eq:epsilonmix}) and (\ref{Eq:M-BFlow} below, is also exhibited.

Even though the differential rate is proportional \cite{spergel12} to $g(\upsilon_{min},\upsilon_E(\alpha))$, for the benefit of the experimentalists, we would like to make more explicit the dependence of the  differential rate on each of the variables entering the expression $g(\upsilon_{min},\upsilon_E(\alpha))$ and in particular to isolate the coefficient of $\cos{\alpha}$ term, which will provide the interesting modulation amplitude and make the time dependence explicit. This approach will be even more useful, when one integrates the differential rate to obtain the total event rate.\\
To this end, we will find it useful to useful to expand $g(\upsilon_{min},\upsilon_E(\alpha))$ in in powers of $\delta$, keeping terms up to linear in $\delta \approx 0.135$. We found it convenient to  express all velocities in units of the Sun's velocity $\upsilon_0$ to  obtain:
\beq
\upsilon_0 g(\upsilon_{min},\upsilon_E(\alpha))=\Psi_0(x)+\Psi_1(x)\cos{\alpha}, \quad x=\frac{\upsilon_{min}}{\upsilon_{0}}
\eeq
$\Psi_0(x)$ represents the quantity relevant for the average rate and $\Psi_1(x)$, which is proportional to $\delta$, represents the effects of modulation. \\
In the case of a M-B distribution these functions take the following form:
\beq
\Psi_0(x)=\frac{1}{2}
  \left [\mbox{erf}(1-x)+\mbox{erf}(x+1)+\mbox{erfc}(1-y_{\mbox{\tiny{esc}}})+\mbox{erfc}(y_{\mbox{\tiny{esc}}}+1)-2 \right ]
  \label{Eq:Psi0MB}
\eeq
\barr
\Psi_1(x)&=&\frac{1}{2} \delta 
   \left[\frac{ -\mbox{erf}(1-x)-\mbox{erf}(x+1)-\mbox{erfc}(1-y_{\mbox{\tiny{esc}}})-
   \mbox{erfc}(y_{\mbox{\tiny{esc}}}+1)}{2} \right . \nonumber\\
  && \left . +\frac{ e^{-(x-1)^2}}{\sqrt{\pi }}
   +\frac{
   e^{-(x+1)^2}}{\sqrt{\pi }}-\frac{ e^{-(y_{\mbox{\tiny{esc}}}-1)^2}}{\sqrt{\pi
   }}-\frac{ e^{-(y_{\mbox{\tiny{esc}}}+1)^2}}{\sqrt{\pi }}+1\right]
   \label{Eq:Psi1MB}
\earr
where erf$(x)$ and erfc$(x)$ are the error function and its complement, respectively, and $y_{esc}=\upsilon_{esc}/\upsilon_0\approx2.84$.
\begin{figure}
\begin{center}
\rotatebox{90}{\hspace{0.0cm} $\Psi_1(x)\rightarrow$}
\includegraphics[height=0.4\textwidth]{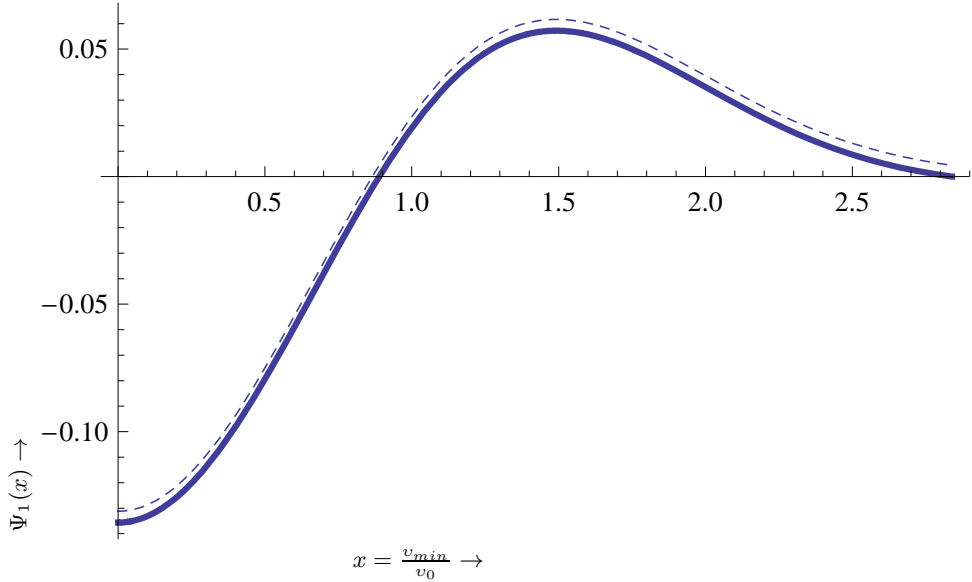}
\\
{\hspace{-2.0cm} $x=\frac{\upsilon_{min}}{\upsilon_{0}}\rightarrow$}
\end{center}
\caption{ The  function $\Psi_1(x),\,x=\upsilon_{min}/\upsilon_0$ , relevant for the modulated differential event rate, evaluated with the standard M-B distribution (dotted line) together with the corresponding M-B case with an upper cut off (solid line). The behavior of $\Psi_0(x)$ is similar.
 \label{Fig:effyesc}}
\end{figure}

 In the case of the flows they were derived from the semi-analytic approximations of simulations as discussed by Kuhlen, Lisanti and Spergel \cite{spergel12}.
 For  isotropic debris flows one finds:
\beq
\Psi_0(x)=\left \{\begin{array}{ll}\frac{1}{y_f},&0<x<y_f-1\\ \frac{1+y_f-x}{2 y_f},&y_f-1<x<1+y_f\\0,&x>1+y_f\end{array}\right . ,\quad y_f=\frac{\upsilon_{flow}}{\upsilon_0}
\label{Eq:Psi0Deb}
\eeq
\beq
\Psi_1(x)=\delta \left \{\begin{array}{ll}0,&0<x<y_f-1\\  \frac{x-y_f}{4 y_f}, &y_f-1<x<1+y_f\\0,&x>1+y_f\end{array}\right . ,\quad y_f=\frac{\upsilon_{flow}}{\upsilon_0}
\label{Eq:Psi1Deb}
\eeq
\\
We note that the variable $x$ depends on the nuclear recoil energy $E_R$ as well as the WIMP-nucleus reduced mass.
 As we shall see below, there is an additional dependence of the rates on $E_R$ coming from the nuclear form factor.

\begin{figure}
\begin{center}
\rotatebox{90}{\hspace{0.0cm} $\epsilon(x)\rightarrow$}
\includegraphics[height=0.4\textwidth]{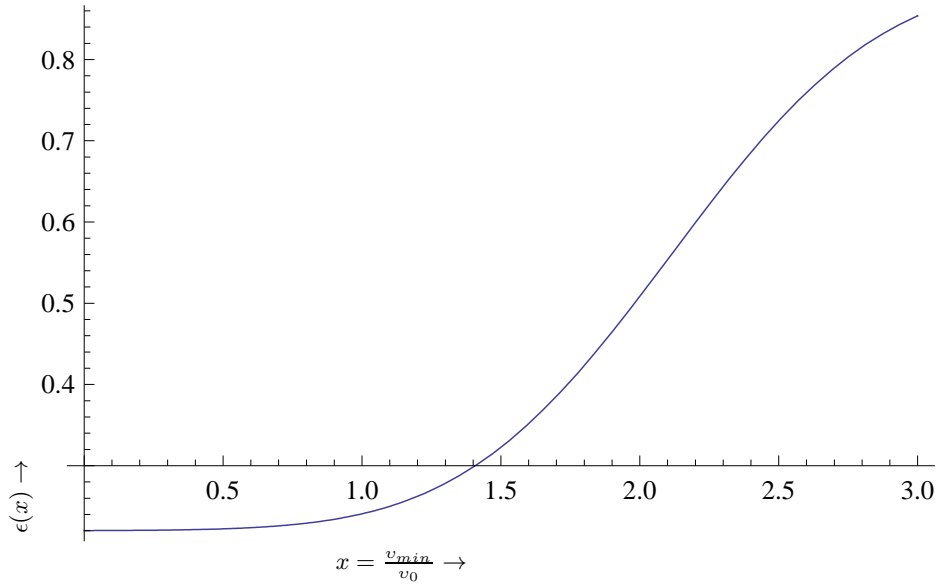}
\\
{\hspace{-2.0cm} $x=\frac{\upsilon_{min}}{\upsilon_{0}}\rightarrow$}
\end{center}
\caption{ The  function $\epsilon(x),\,x=\upsilon_{min}/\upsilon_0$ as a function of $x$, which gives a possible combination of a M-B distribution and  debris flows \cite{spergel12}.
 \label{fig:epsilon}}
\end{figure}

\begin{figure}
\begin{center}
\subfloat[]
{
\rotatebox{90}{\hspace{0.0cm} $\Psi_0(x)\rightarrow$}
\includegraphics[height=0.27\textwidth]{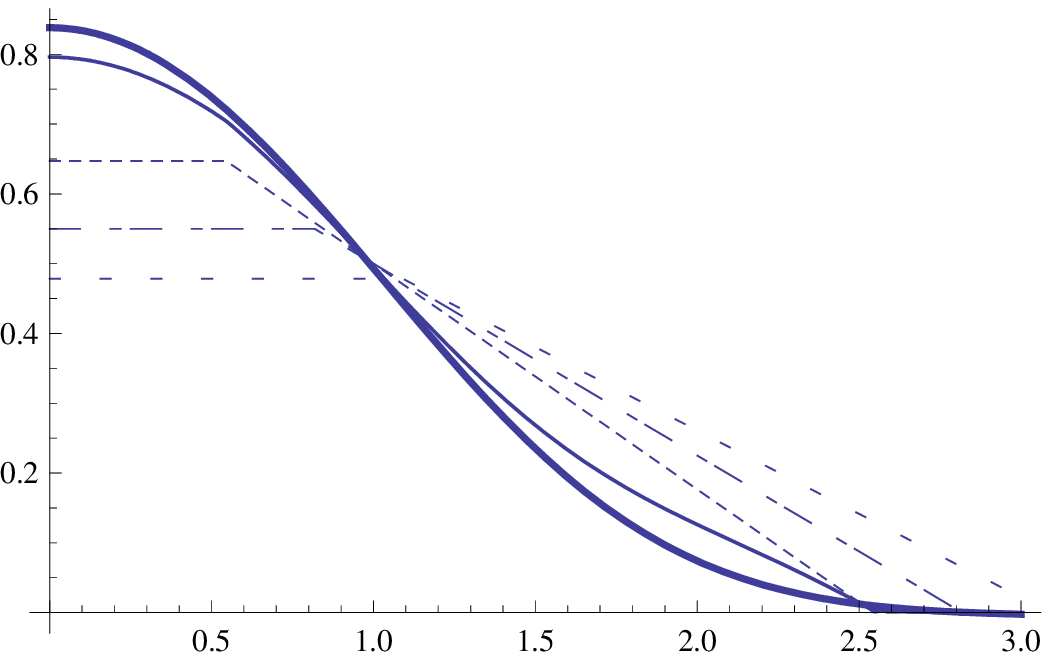}
}
\subfloat[]
{
\rotatebox{90}{\hspace{0.0cm} $\Psi_1(x)\rightarrow$}
\includegraphics[height=0.27\textwidth]{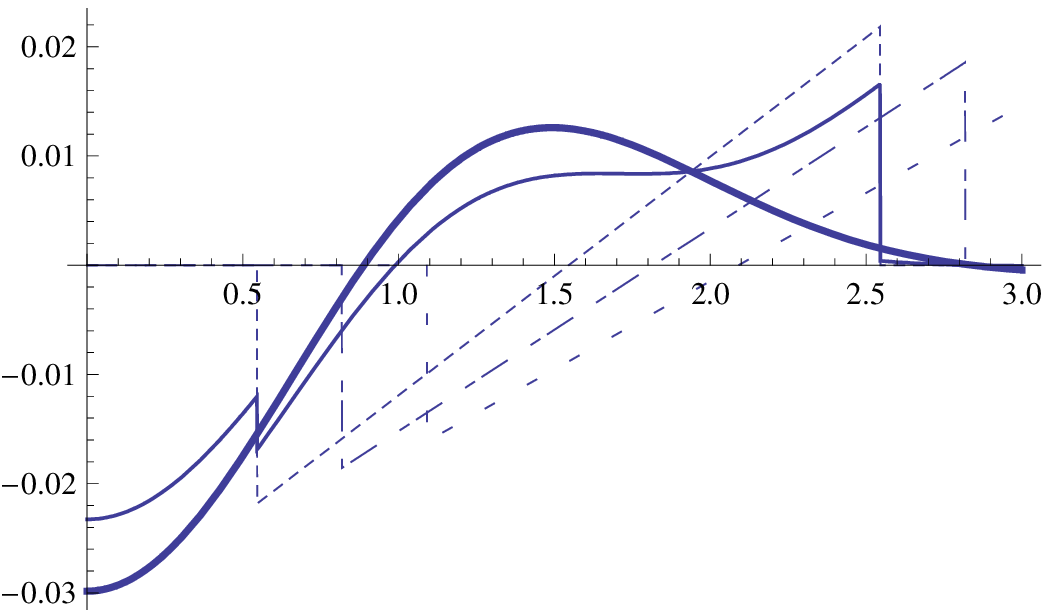}
}
\\
{\hspace{-2.0cm} $x=\frac{\upsilon_{min}}{\upsilon_{0}}\rightarrow$}
\end{center}
\caption{ The  functions $\Psi_0(x)$ and $\Psi_1(x)$ as  functions of $x={\upsilon_{min}}/{\upsilon_{0}}$, given by Eqs (\ref{Eq:Psi0MB})-(\ref{Eq:Psi1Deb}) and (\ref{Eq:M-BFlow}). Note also that the variable $x$ depends on the nuclear recoil energy $E_R$ as well as the WIMP-nucleus reduced mass. Otherwise the labeling of the curves is the same as that of Fig. \ref{fig:flowv}.
 \label{fig:Psi01}}
\end{figure}

At Earth-frame velocities greater than 450 km/s,
debris flow comprises more than half of the dark matter at the Sun's location, and up to
$80\%$ at even higher velocities \cite{spergel12}. The combination of debris flows and standard M-B 
is provided by  the relative density fraction $\epsilon$ of the Via Lactea 2 particles tagged as debris flows 
in the radial shell $7.5 \mbox{ kpc} < r < 9.5$ kpc compared to the total number of particles in the simulation from 7.5 to 9.5 kpc. In the VL2 simulation it is well fitted \cite{spergel12} by a Gauss error function: 
\beq
\epsilon(x)=0.22 + 0.34 \left (\mbox{erf}\left (x \frac{220}{185}-\frac{465}{185}\right ) + 1\right )
\label{Eq:epsilonmix}
\eeq
This function is exhibited in Fig. \ref{fig:epsilon}.
In this case we find:
\beq
\Psi_i(x)\rightarrow\left [1-\epsilon(x)\right ]\Psi^{MB}_i(x)+\epsilon(x)\Psi^{f}_i(x),\quad i=0,1
\label{Eq:M-BFlow}
\eeq
The behavior of the functions $\Psi_0(x)$ and $\Psi_1(x)$ is exhibited in Fig. \ref{fig:Psi01}. As expected, in the case of the flows $\Psi_0(x)$ falls off linearly for large values of $x$, but $\Psi_1(x)$ increases linearly. Note, however,  that in all cases  $\Psi_1(x)$  takes both positive and negative values, which affects the location of the maximum of the modulated rate as a function of time,   obtained after $\Psi_1(x)$ is multiplied by $\cos{\alpha}$. The location of the maximum depends on the target and the WIMP mass as we will see below. 

\begin{figure}
\begin{center}
\rotatebox{90}{\hspace{1.0cm} $F^2(\frac{E_R}{Q_0(A)})\rightarrow$}
\includegraphics[width=0.8\textwidth]{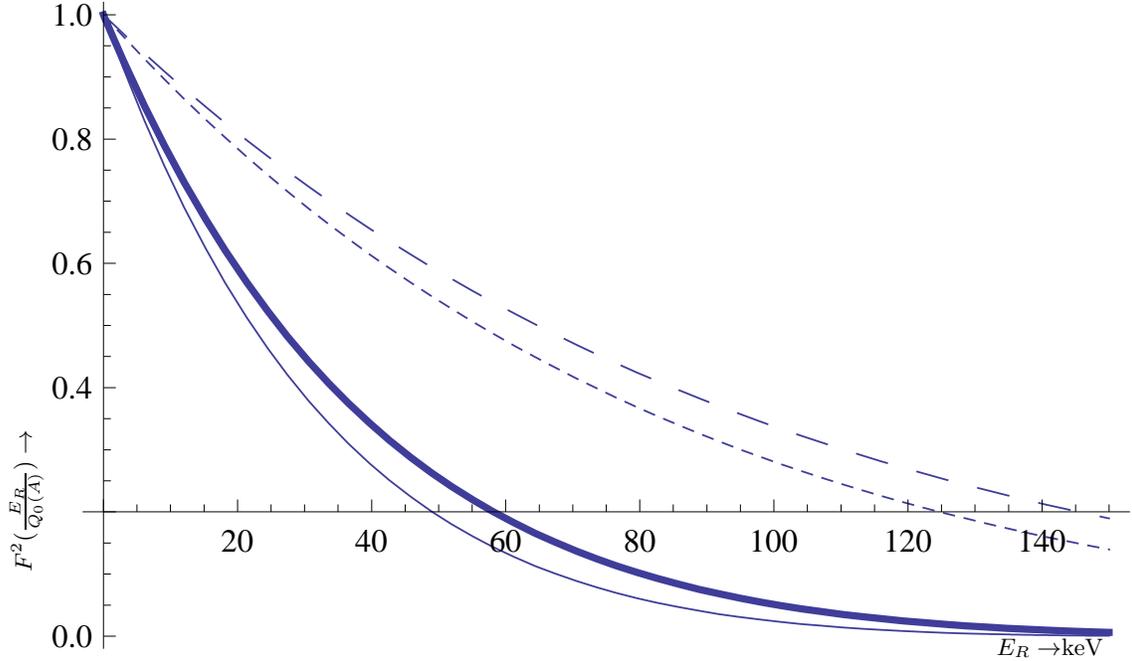}
{\hspace{-2.0cm} $E_R\rightarrow$keV}
\end{center}
\caption{ The square of the nuclear form factor $F(\frac{E_R}{Q_0(A)})$, entering Eqs (\ref{Eq:ddtdydhdu}) and (\ref{defineu}),   for a heavy target, e.g. $^{127}$I, (thin solid curve) and an intermediate weight target , e.g. $^{73}$Ge,  (short-dashed curve) obtained in the context of the shell model employed in this work. For comparison  we show the same quantities using the Helm type form factor, indicated by the thick solid and long-dashed curves for $^{127}$I and $^{73}$Ge respectively. For lighter targets the effect of the form factor is small.
 \label{fig:FFsq}}
\end{figure}

Once these functions are known, the formalism to obtain the direct detection rates is fairly well known (see e.g. the recent reviews \cite{JDV06a,VerMou11}). So, we will briefly discuss its essential elements here.
The differential event rate can be cast in the form:
\beq
\left . \frac{d R}{ d E_R}\right |_A=\left . \frac{dR_0}{dE_R}\right |_A+\left . \frac{d{\tilde H}}{dE_R}\right |_A \cos{\alpha}
\eeq
where the first term represents the time averaged (non modulated) differential event rate, while the second  gives the time dependent (modulated) one due to the motion of the Earth (see below). Furthermore one finds
\barr
\left .\frac{d R_0}{ d E_R}\right |_A&=&\frac{\rho_{\chi}}{m_{\chi}}\frac{m_t}{A m_p} \sigma_n\left (\frac{\mu_r}{\mu_p} \right )^2 \sqrt{<\upsilon^2>} A^2\frac{1}{Q_0(A)}\frac{d t}{du},\nonumber\\
\left . \frac{d {\tilde H}}{ d E_R}\right |_A&=&\frac{\rho_{\chi}}{m_{\chi}}\frac{m_t}{A m_p} \sigma_n\left (\frac{\mu_r}{\mu_p} \right )^2 \sqrt{\langle \upsilon^2\rangle } A^2 \frac{1}{Q_0(A)} \frac{d h}{du}
\label{drdu}
\earr
with  $\mu_r$ ($\mu_p$) the WIMP-nucleus (nucleon) reduced mass, $A$  the nuclear mass number, $\sigma_n$  the elementary WIMP-nucleon cross section, $ m_{\chi}$  the WIMP mass and $m_t$ the mass of the target. 
Furthermore one can show that
\beq
\frac{d t}{d u}=\sqrt{\frac{2}{3}} a^2 F^2(u)   \Psi_0(a \sqrt{u}),\quad \frac{d h}{d u}=\sqrt{\frac{2}{3}} a^2 F^2(u) \Psi_1(a \sqrt{u}).
\label{Eq:ddtdydhdu}
\eeq
Here $a=(\sqrt{2} \mu_r b \upsilon_0)^{-1}$ with  $\upsilon_0$ the velocity of the Sun around the center of the Galaxy and $b$ the nuclear harmonic oscillator size parameter characterizing the nuclear wave function.  $ u$ is the energy transfer $E_R$ in dimensionless units given by
\begin{equation}
 u=\frac{E_R}{Q_0(A)}~~,~~Q_{0}(A)=[m_pAb^2]^{-1}=60A^{-4/3}\mbox{ MeV}
\label{defineu}
\end{equation}
and $F(u)$ is the nuclear form factor. In the present calculation the form factors were obtained in the context of the nuclear shell model in the spirit of Ref. \cite{DIVA00,Ressa,Ressb} with suitably adjusted size parameter $b$. We will compare them, however, with the phenomenological  Helm type form factors \cite{DKG07}, preferred by the experimentalists (for a recent discussion of the various types of form factors see Ref. \cite{Cannoni11}). Anyway, the form factor is important in the case of a large reduced mass, i.e. when large recoil energies are possible (see Fig. \ref{fig:FFsq}). 
Note that the parameter $a$ depends  on the WIMP mass, the target and the velocity distribution.\\
Sometimes one  writes the differential rate as:
\beq
\left .\frac{d R}{ d E_R}\right |_A=\frac{\rho_{\chi}}{m_{\chi}}\frac{m_t}{A m_p} \sigma_n \left ( \frac{\mu_r}{\mu_p} \right )^2 \sqrt{\langle\upsilon^2\rangle} A^2 \frac{1}{Q_0(A)}\left(\frac{d t}{du}(1+ H(a \sqrt{E_R/Q_0(A)}) \cos{\alpha}\right )
\label{dhduH}
\eeq
In this formulation $H(a \sqrt{E_R/Q_0(A)}) $, the ratio of the modulated to the non modulated differential rate, gives the relative differential modulation amplitude. It coincides with the ratio $\Psi_1(a \sqrt{E_R/Q_0(A)})/\Psi_0(a \sqrt{E_R/Q_0(A)})$, i.e. it is independent of the nuclear form factor and  depends only on the reduced mass and the velocity distribution. It is, thus, the same for both the coherent and the spin mode. Note that it can take both positive and negative values, which affects the location of the maximum of the modulated rate as a function of $\alpha$. For the convenience of the analysis of experiments, however, we will present our results in the form of Eq. (\ref{drdu}).

Sometimes, as is the case for the DAMA experiment, the target has many components. In such cases the above formalism can be applied as follows:
\beq
\left .\frac{dR}{dE_R}\right |_A\rightarrow \sum_i X_i\left . \frac{dR}{dE_R}\right |_{A_i},\quad u\rightarrow u_i,\quad X_i=\mbox{the fraction of the component } A_i\mbox{ in the target}
\eeq
We will not, however, pursue such an analysis.

\section{Some results on differential rates}
\begin{figure}
\begin{center}
\subfloat[]
{
\rotatebox{90}{\hspace{0.0cm} $dR/dQ\rightarrow$kg/(y keV)}
\includegraphics[height=.17\textheight]{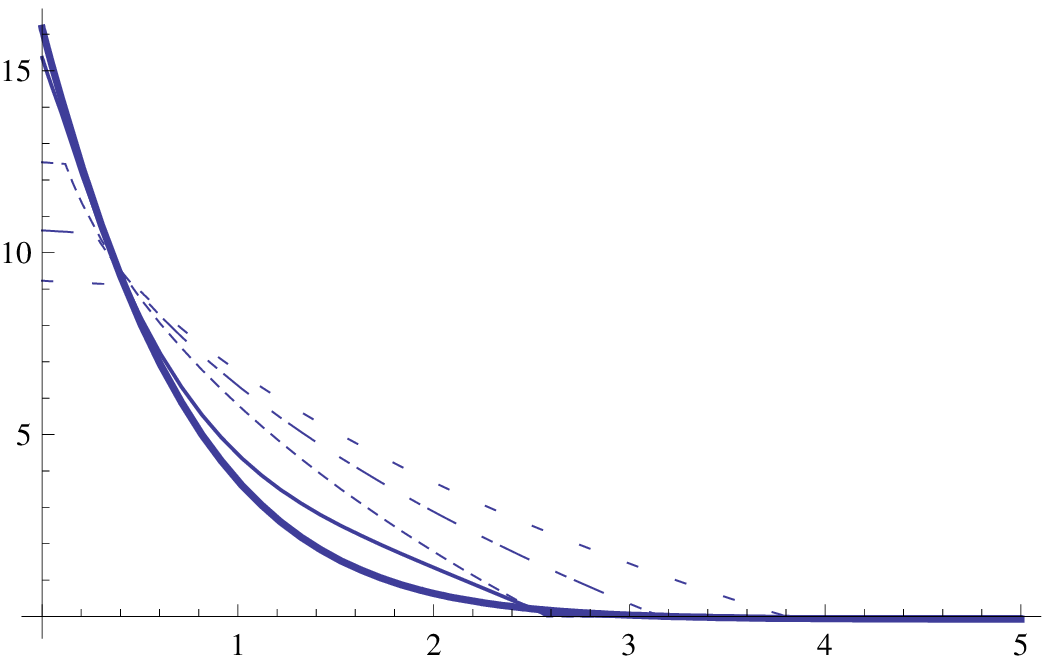}
}
\subfloat[]
{
\rotatebox{90}{\hspace{0.0cm} $dR/dQ\rightarrow$kg/(y keV)}
\includegraphics[height=.17\textheight]{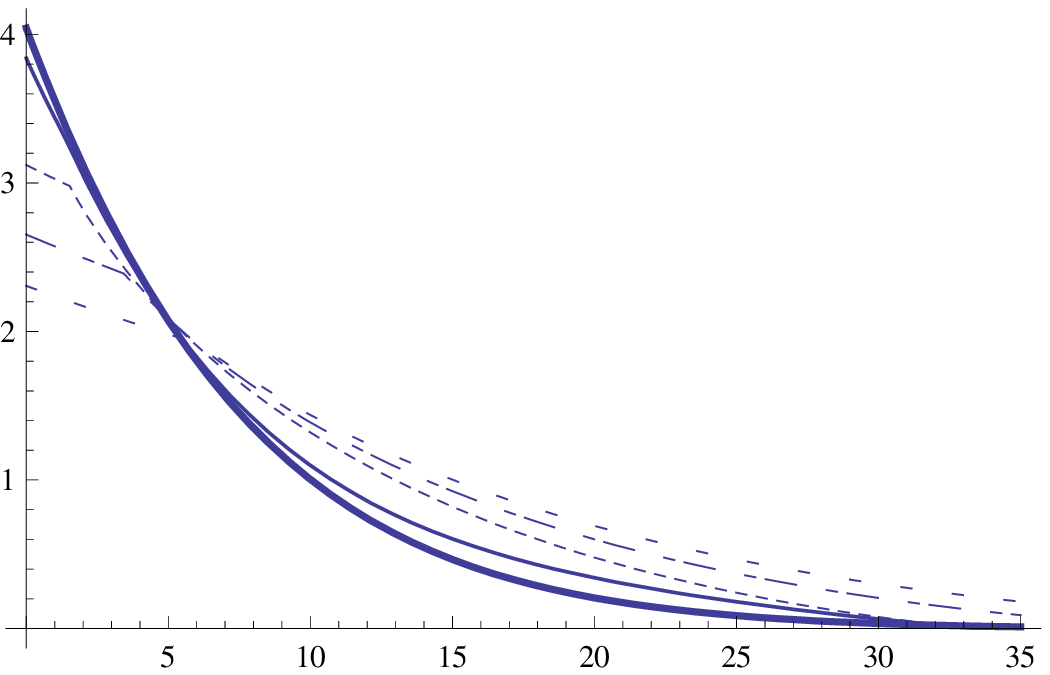}
}
\\
\subfloat[]
{
\rotatebox{90}{\hspace{0.0cm} $dR/dQ\rightarrow$kg/(y keV)}
\includegraphics[height=.17\textheight]{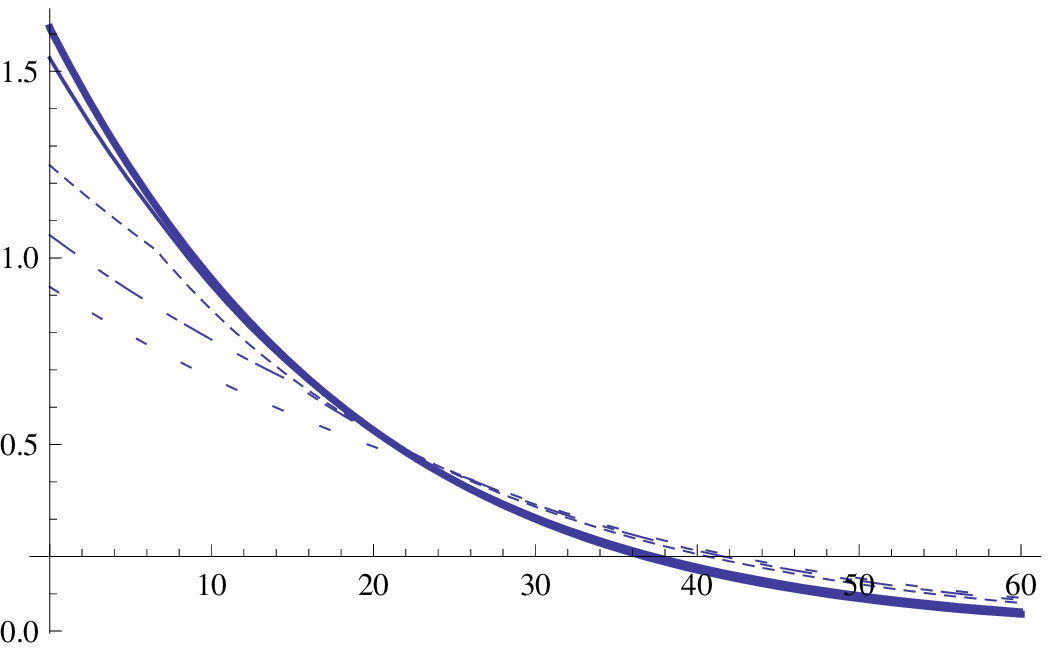}
}
\subfloat[]
{
\rotatebox{90}{\hspace{0.0cm} $dR/dQ\rightarrow$kg/(y keV)}
\includegraphics[height=.17\textheight]{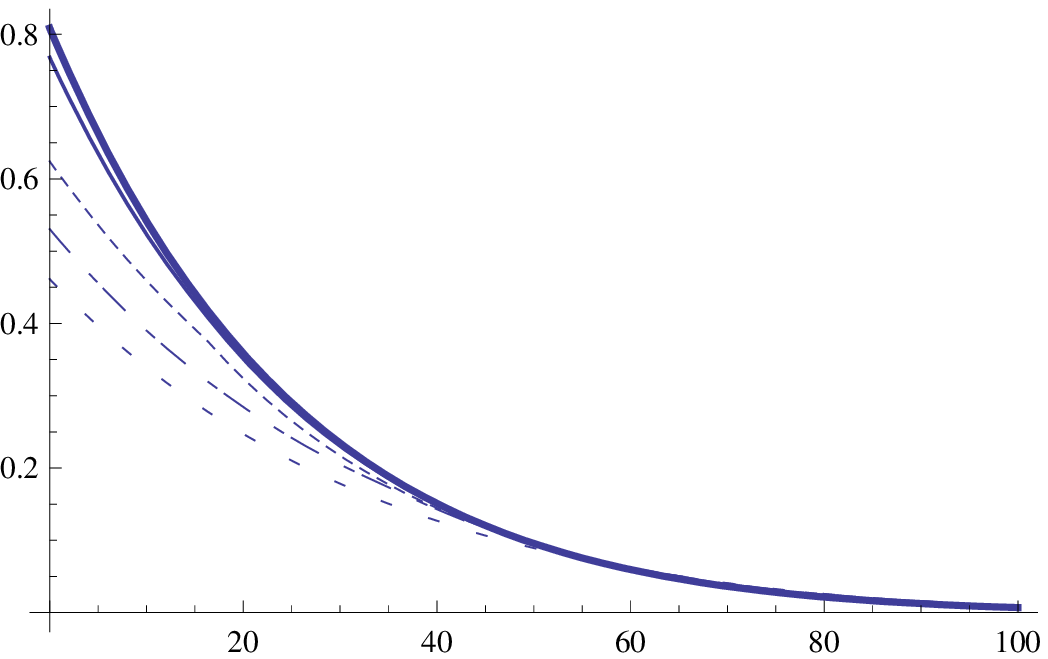}
}
\\
{\hspace{-2.0cm} $Q\rightarrow$keV}
\end{center}
\caption{ The differential rate $\frac{dR}{dQ}$,   as a function of the recoil energy for a heavy target, e.g. $^{127}$I assuming a nucleon cross section of $10^{-7}$pb. Panels (a) (b), (c) and (d) correspond to to 5, 20, 50 and 100 GeV WIMP masses. Otherwise the notation is the same as that of Fig. \ref{fig:flowv}.
 \label{fig:dRdQ127}}
\end{figure}

\begin{figure}
\begin{center}
\subfloat[]
{
\rotatebox{90}{\hspace{0.0cm} $d\tilde{H}/dQ\rightarrow$kg/(y keV)}
\includegraphics[height=.17\textheight]{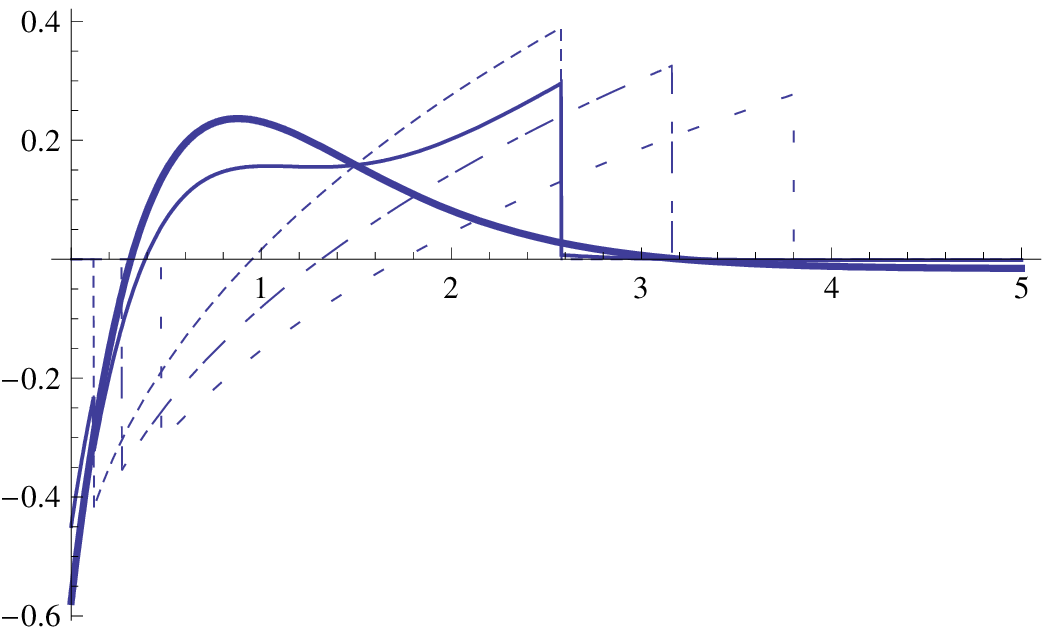}
}
\subfloat[]
{
\rotatebox{90}{\hspace{0.0cm} $d{\tilde H}/dQ\rightarrow$kg/(y keV)}
\includegraphics[height=.17\textheight]{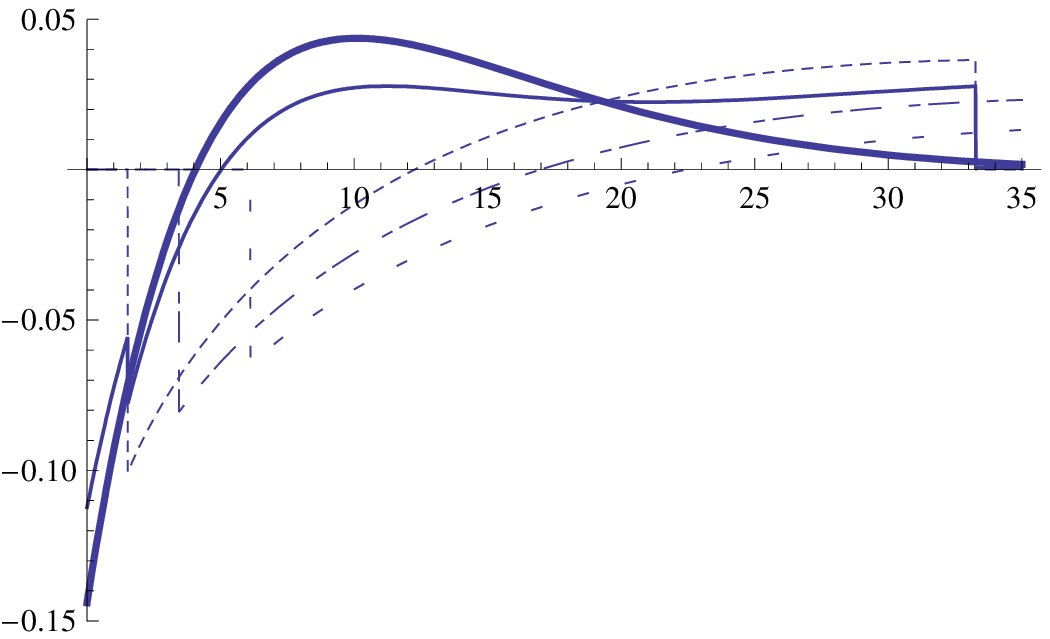}
}
\\
\subfloat[]
{
\rotatebox{90}{\hspace{0.0cm} $d\tilde{H}/dQ\rightarrow$kg/(y keV)}
\includegraphics[height=.17\textheight]{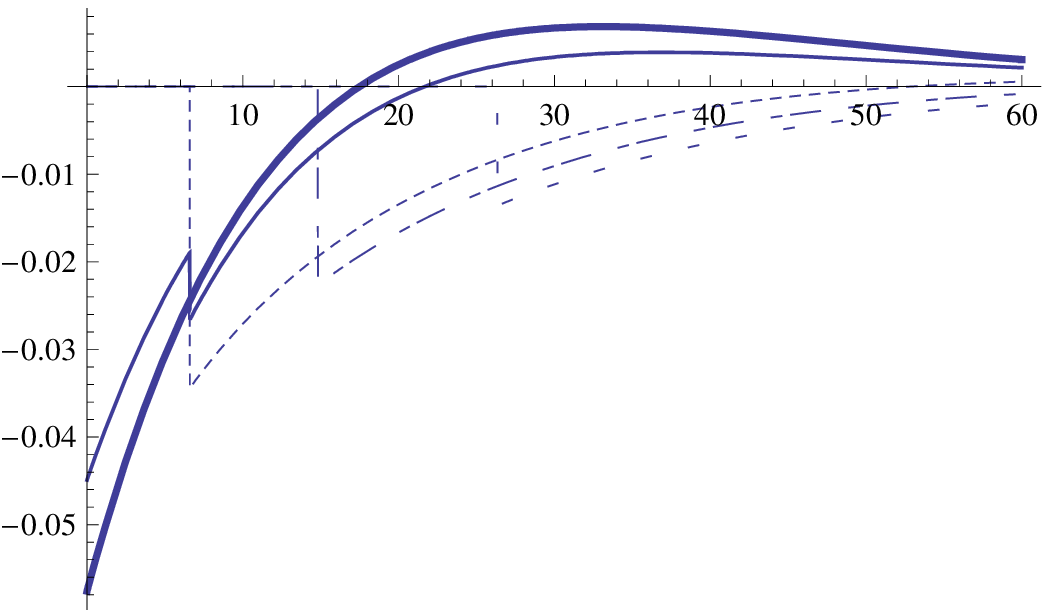}
}
\subfloat[]
{
\rotatebox{90}{\hspace{0.0cm} $d{\tilde H}/dQ\rightarrow$kg/(y keV)}
\includegraphics[height=.17\textheight]{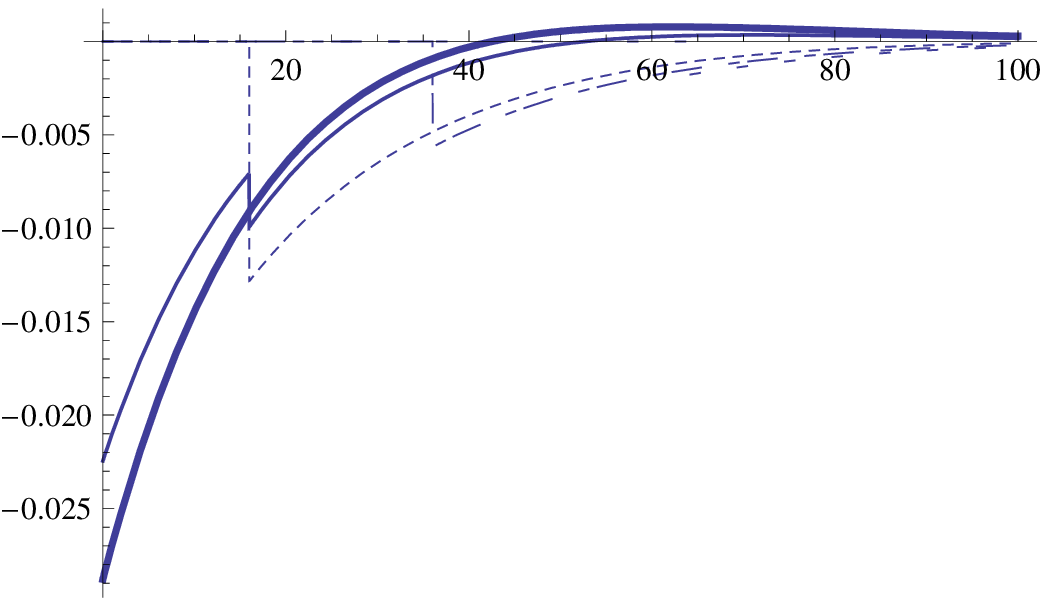}
}
\\
{\hspace{-2.0cm} $Q\rightarrow$keV}
\end{center}
\caption{ The differential rate $\frac{d\tilde{H}}{dQ}$,   as a function of the recoil energy for a heavy target, e.g. $^{127}$I assuming a nucleon cross section of $10^{-7}$pb. Panels (a) (b), (c) and (d) correspond to to 5, 20, 50 and 100 GeV WIMP masses. Otherwise the notation is the same as that of Fig. \ref{fig:flowv}.
 \label{fig:dHdQ127}}
\end{figure}

\begin{figure}
\begin{center}
\subfloat[]
{
\rotatebox{90}{\hspace{0.0cm} $dR/dQ\rightarrow$kg/(y keV)}
\includegraphics[height=.17\textheight]{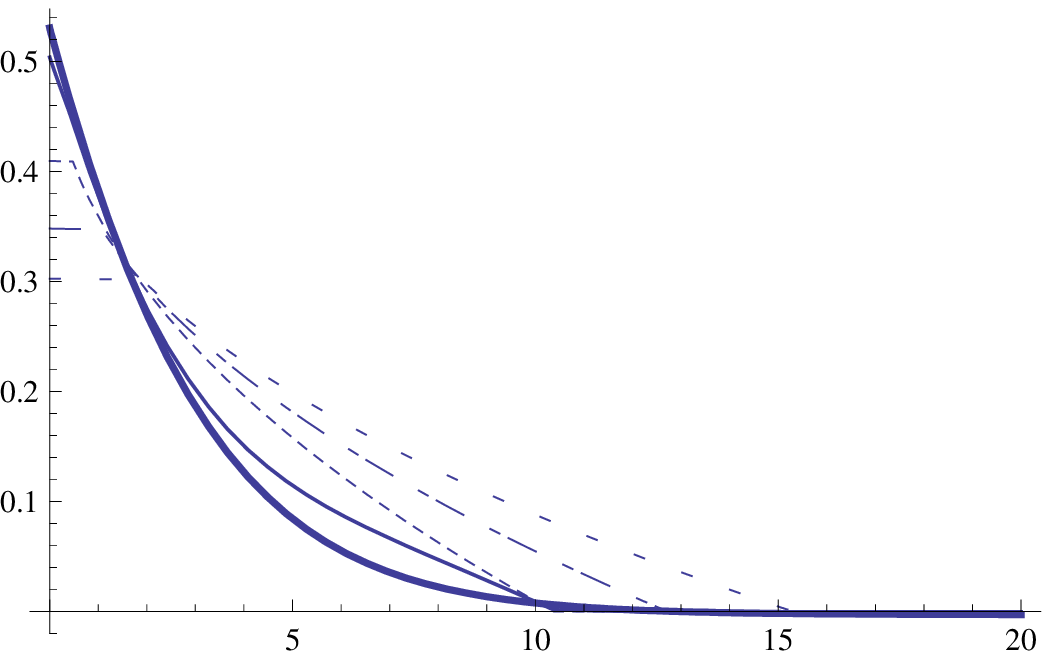}
}
\subfloat[]
{
\rotatebox{90}{\hspace{0.0cm} $dR/dQ\rightarrow$kg/(y keV)}
\includegraphics[height=.17\textheight]{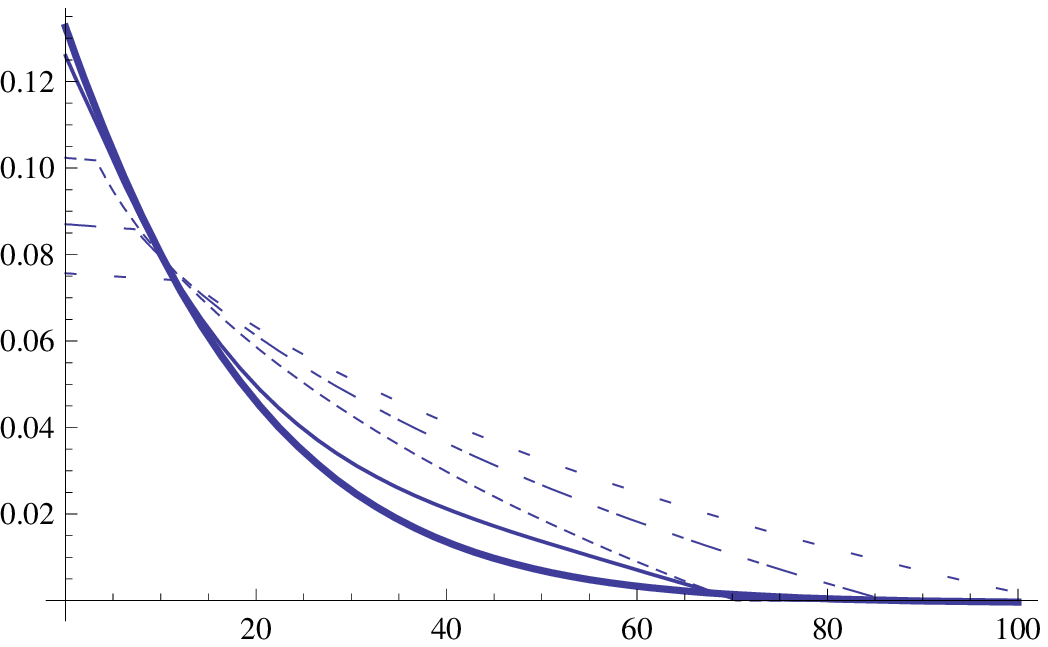}
}
\\
\subfloat[]
{
\rotatebox{90}{\hspace{0.0cm} $dR/dQ\rightarrow$kg/(y keV)}
\includegraphics[height=.17\textheight]{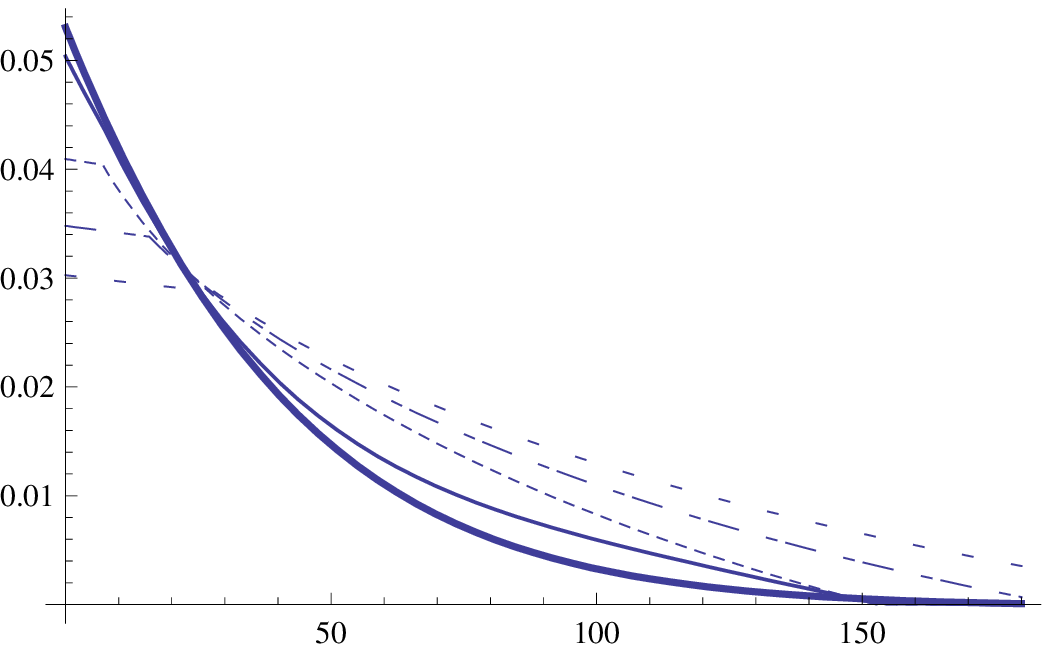}
}
\subfloat[]
{
\rotatebox{90}{\hspace{0.0cm} $dR/dQ\rightarrow$kg/(y keV)}
\includegraphics[height=.17\textheight]{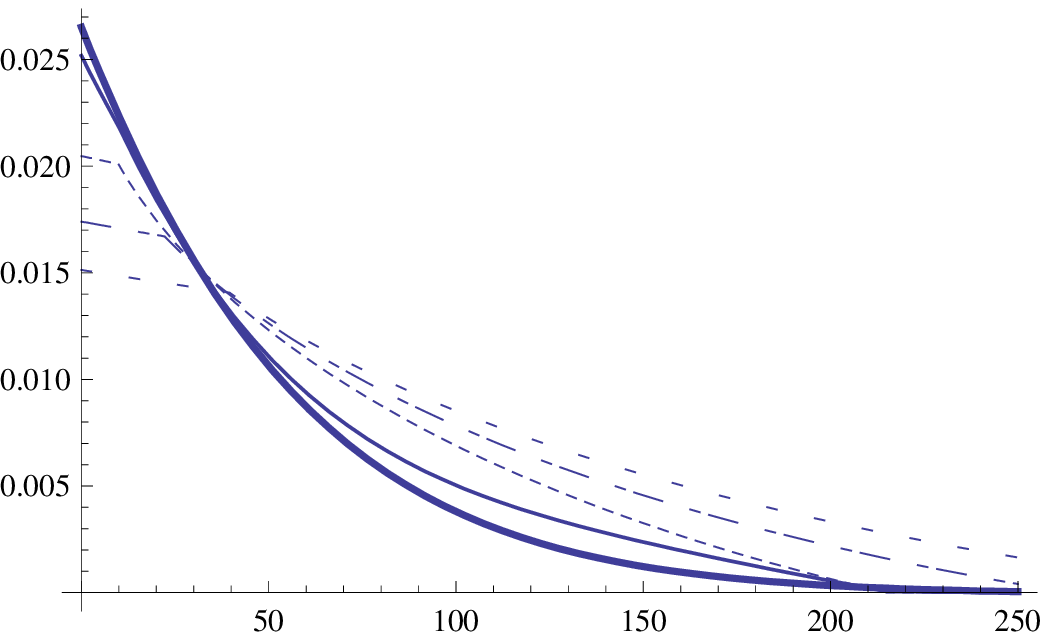}
}
\\
{\hspace{-2.0cm} $Q\rightarrow$keV}
\end{center}
\caption{ The differential rate $\frac{dR}{dQ}$,   as a function of the recoil energy for a light target, e.g. $^{23}$Na assuming a nucleon cross section of $10^{-7}$pb. Panels (a) (b), (c) and (d) correspond to to 5, 20, 50 and 100 GeV WIMP masses. Otherwise the notation is the same as that of Fig. \ref{fig:flowv}.
 \label{fig:dRdQ23}}
\end{figure}

\begin{figure}
\begin{center}
\subfloat[]
{
\rotatebox{90}{\hspace{0.0cm} $d\tilde{H}/dQ\rightarrow$kg/(y keV)}
\includegraphics[height=.17\textheight]{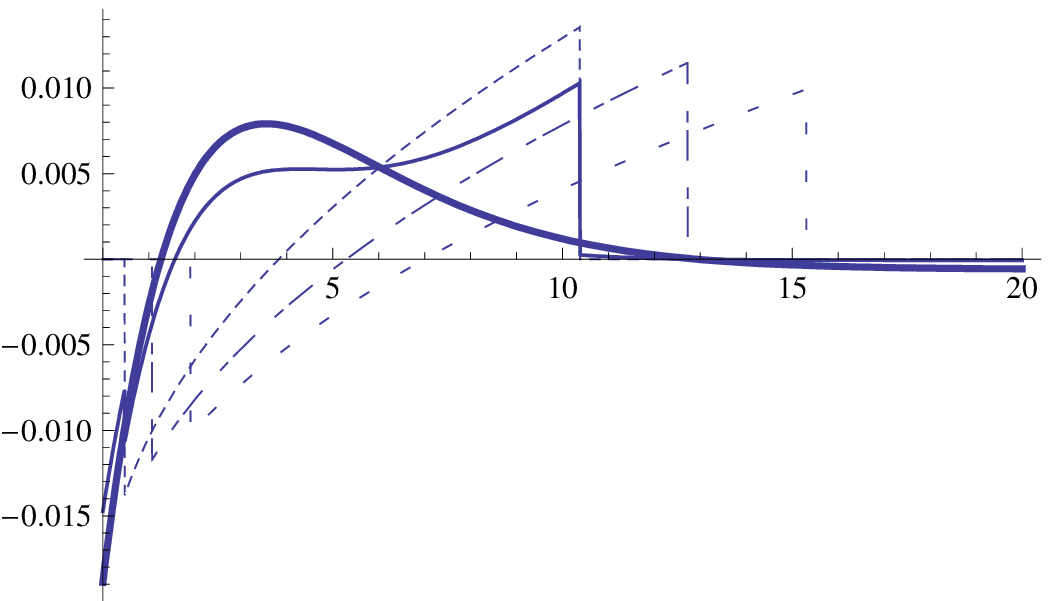}
}
\subfloat[]
{
\rotatebox{90}{\hspace{0.0cm} $d{\tilde H}/dQ\rightarrow$kg/(y keV)}
\includegraphics[height=.17\textheight]{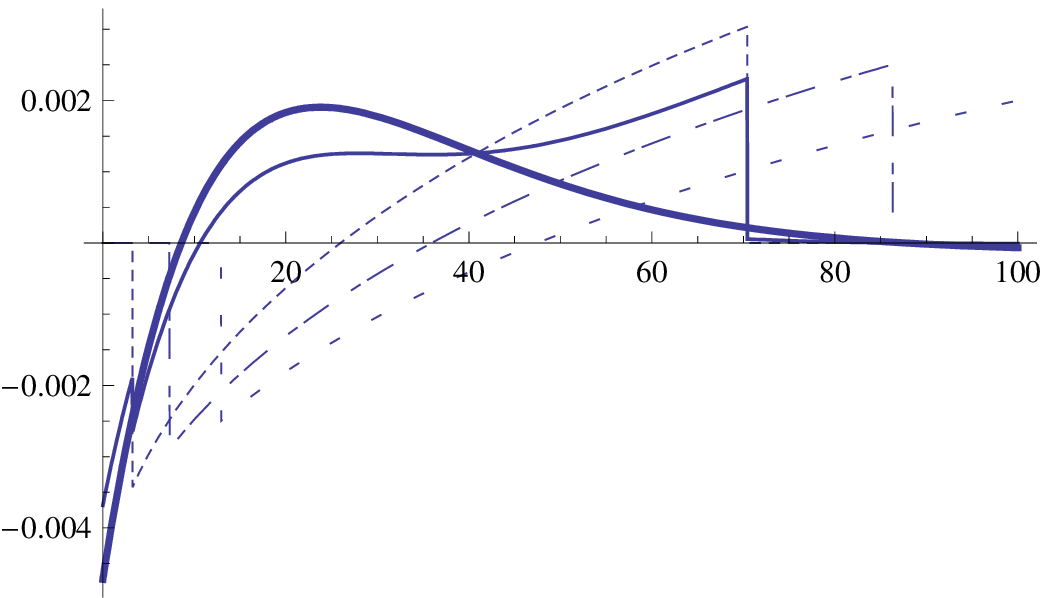}
}
\\
\subfloat[]
{
\rotatebox{90}{\hspace{0.0cm} $d\tilde{H}/dQ\rightarrow$kg/(y keV)}
\includegraphics[height=.17\textheight]{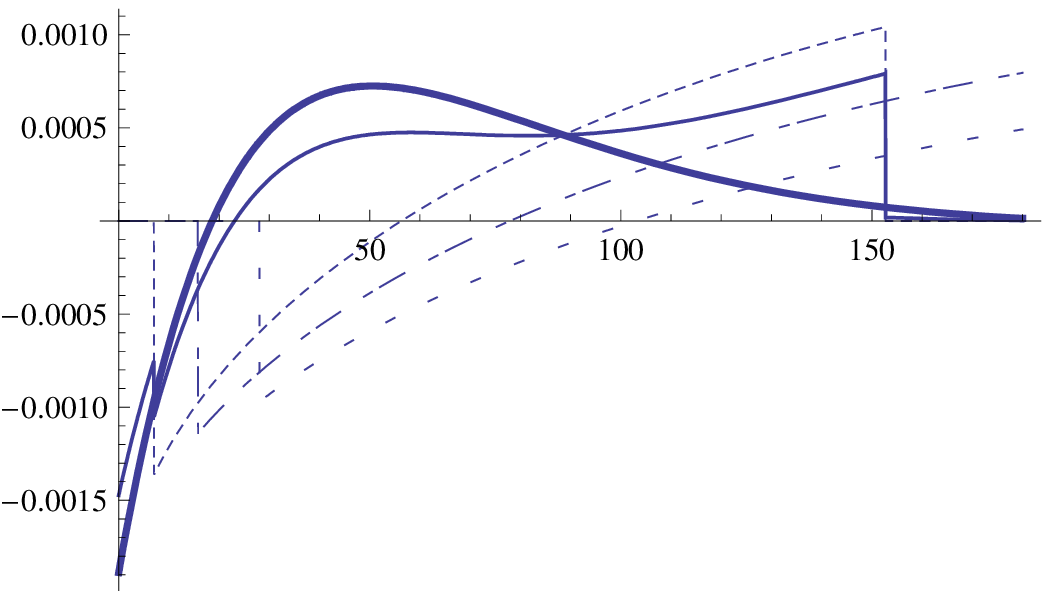}
}
\subfloat[]
{
\rotatebox{90}{\hspace{0.0cm} $d{\tilde H}/dQ\rightarrow$kg/(y keV)}
\includegraphics[height=.17\textheight]{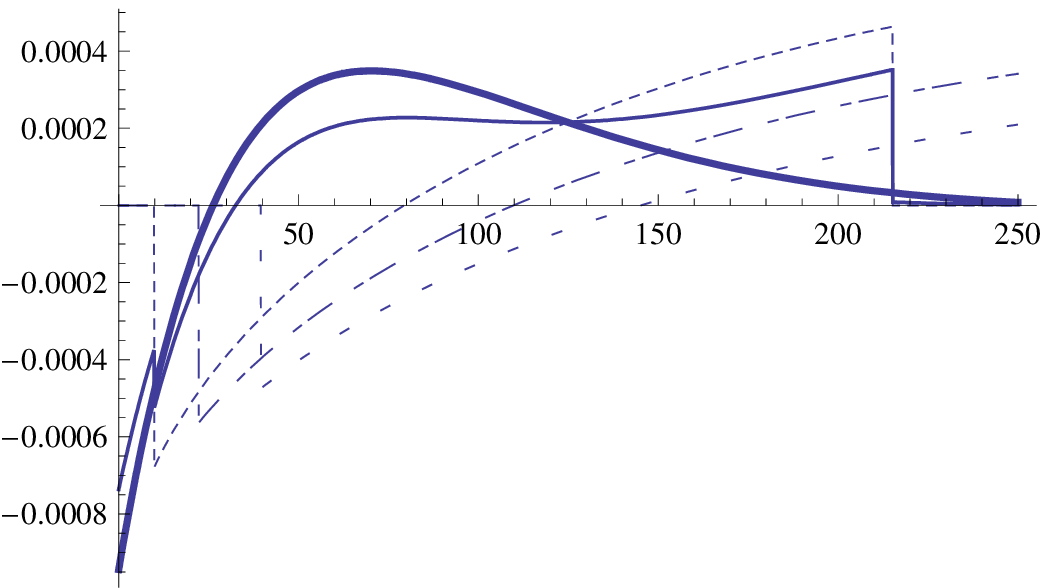}
}
\\
{\hspace{-2.0cm} $Q\rightarrow$keV}
\end{center}
\caption{ The differential rate $\frac{d\tilde{H}}{dQ}$,   as a function of the recoil energy for a light target, e.g. $^{23}$Na assuming a nucleon cross section of $10^{-7}$pb. Panels (a) (b), (c) and (d) correspond to to 5, 20, 50 and 100 GeV WIMP masses. Otherwise the notation is the same as that of Fig. \ref{fig:flowv}.
 \label{fig:dHdQ23}}
\end{figure}

\begin{figure}
\begin{center}
\subfloat[]
{
\rotatebox{90}{\hspace{0.0cm} $dR/dQ\rightarrow$kg/(y keV)}
\includegraphics[height=.17\textheight]{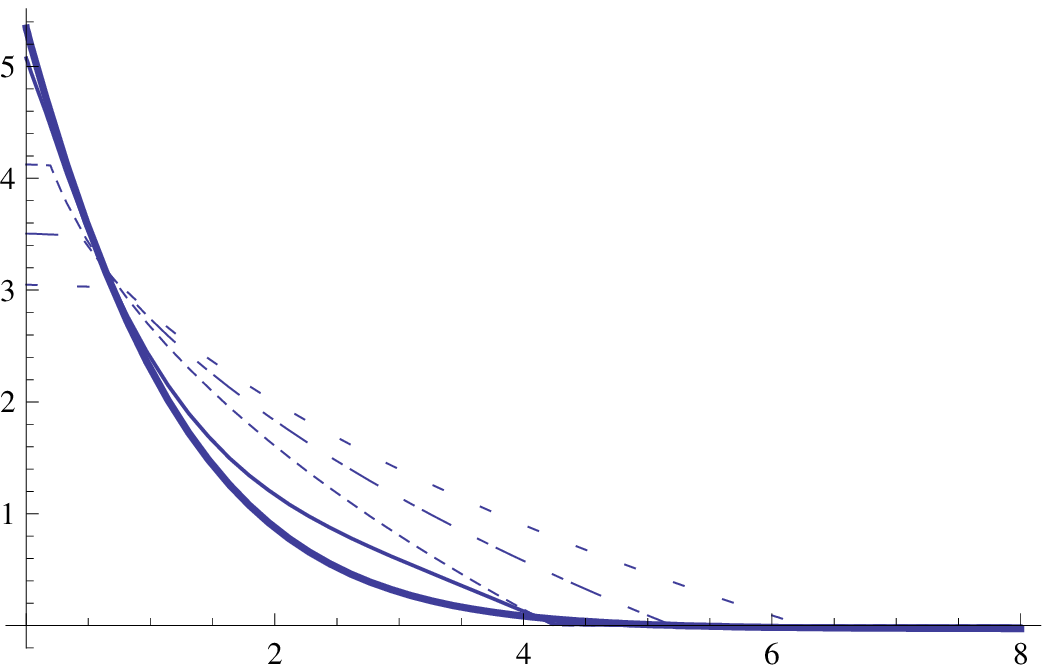}
}
\subfloat[]
{
\rotatebox{90}{\hspace{0.0cm} $dR/dQ\rightarrow$kg/(y keV)}
\includegraphics[height=.17\textheight]{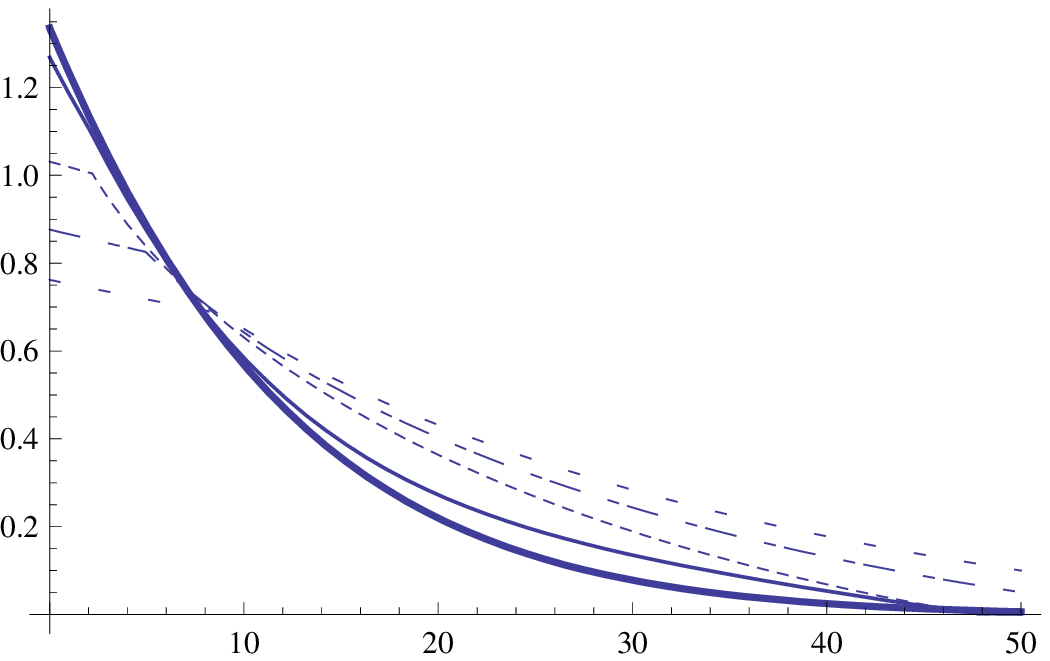}
}
\\
\subfloat[]
{
\rotatebox{90}{\hspace{0.0cm} $dR/dQ\rightarrow$kg/(y keV)}
\includegraphics[height=.17\textheight]{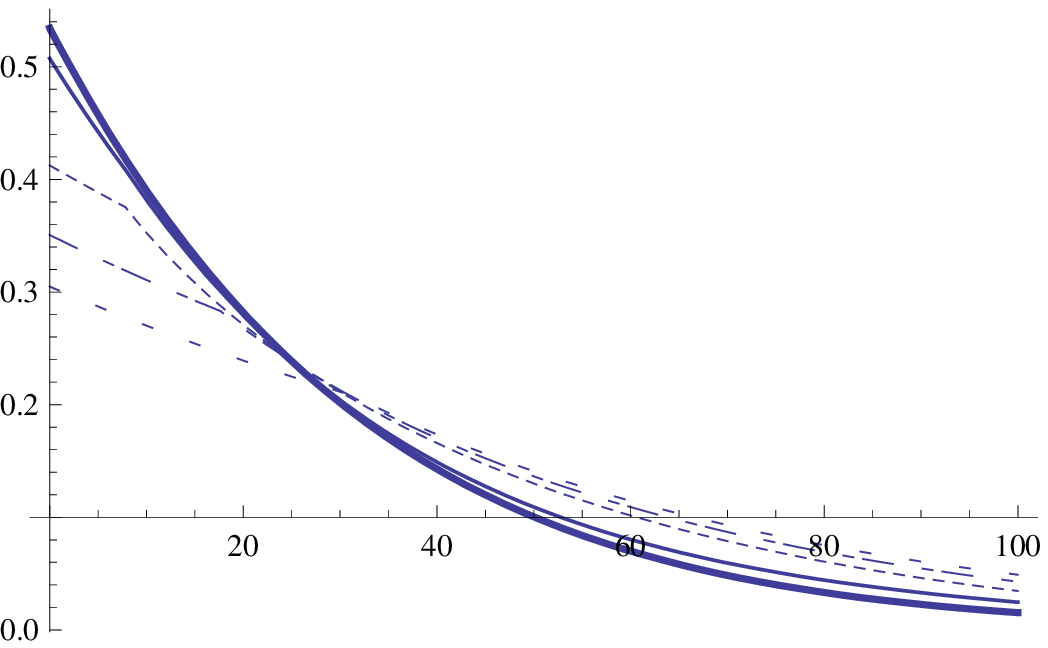}
}
\subfloat[]
{
\rotatebox{90}{\hspace{0.0cm} $dR/dQ\rightarrow$kg/(y keV)}
\includegraphics[height=.17\textheight]{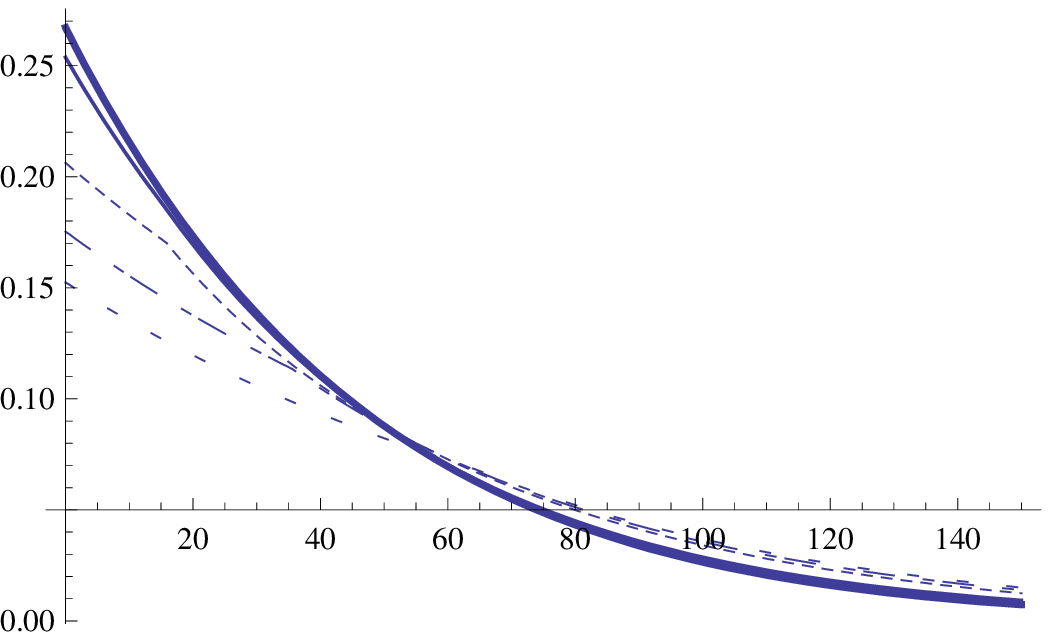}
}
\\
{\hspace{-2.0cm} $Q\rightarrow$keV}
\end{center}
\caption{ The differential rate $\frac{dR}{dQ}$,   as a function of the recoil energy for an intermediate   target, e.g. $^{73}$Ge assuming a nucleon cross section of $10^{-7}$pb. Panels (a) (b), (c) and (d) correspond to to 5, 20, 50 and 100 GeV WIMP masses. Otherwise the notation is the same as that of Fig. \ref{fig:flowv}.
 \label{fig:dRdQ73}}
\end{figure}

We will apply the above formalism in the case of I and Na, which are components of the   NaI detector used in the DAMA experiment \cite{DAMA1,DAMA11} and Ge employed, e.g, by the CoGeNT experiment \cite{CoGeNT11}. The results for the Xe target \cite{XENON10}, \cite{XENON100.11} are similar to those for $^{127}$I, while those  for the $^{19}$F target \cite{PICASSO09,PICASSO11} are similar to those for $^{23}$Na. 
The differential rates $\frac{dR}{dQ}|_A$ and  $\frac{d\tilde{H}}{dQ}|_A$, for $A=127$, $A=23$ and  $A=73$  are exhibited in Figs. \ref{fig:dRdQ127}-\ref{fig:dRdQ73}. 
The nuclear form factor has been included (for a heavy target, like  $^{127}$I or $^{131}$Xe, its effect is sizable even for an energy transfer  of 10 keV, see Fig. \ref{fig:FFsq} and Ref. \cite{JDV12n}).

\begin{figure}
\begin{center}
\subfloat[]
{
\rotatebox{90}{\hspace{0.0cm} $d\tilde{H}/dQ\rightarrow$kg/(y keV)}
\includegraphics[height=.17\textheight]{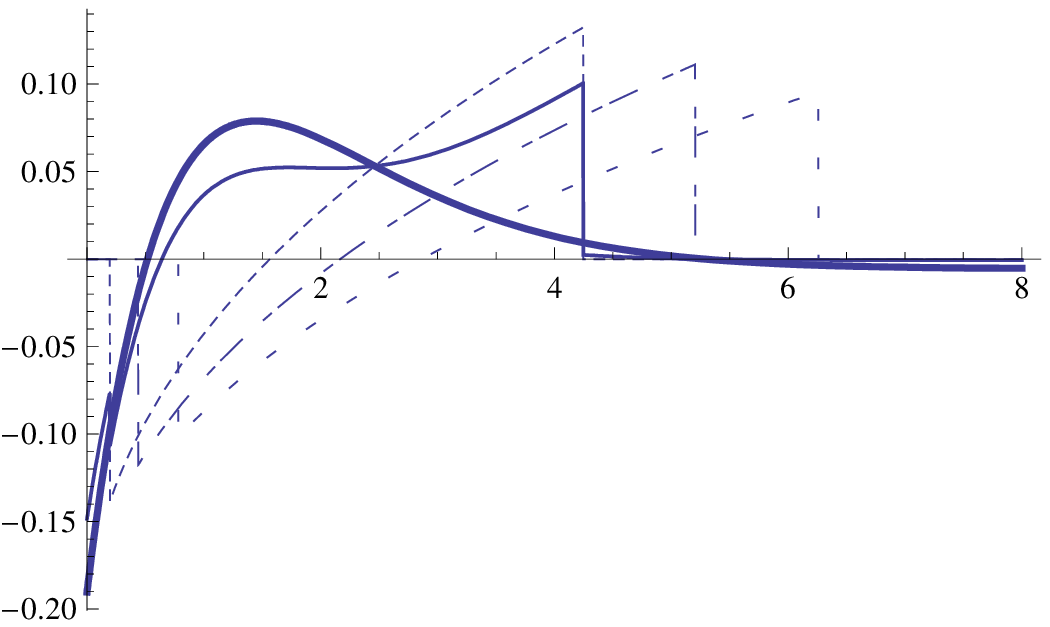}
}
\subfloat[]
{
\rotatebox{90}{\hspace{0.0cm} $d{\tilde H}/dQ\rightarrow$kg/(y keV)}
\includegraphics[height=.17\textheight]{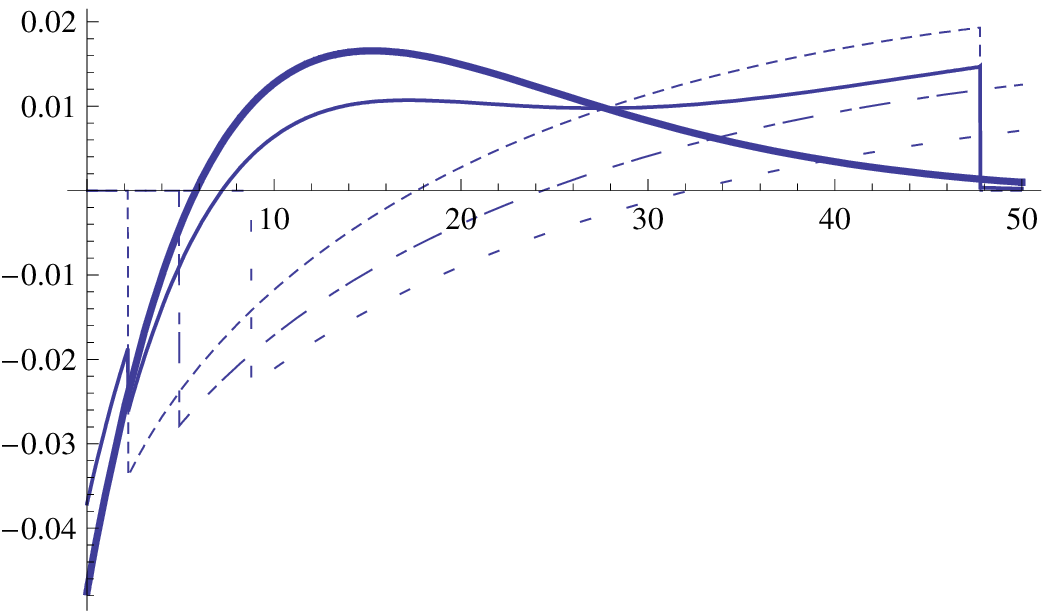}
}
\\
\subfloat[]
{
\rotatebox{90}{\hspace{0.0cm} $d\tilde{H}/dQ\rightarrow$kg/(y keV)}
\includegraphics[height=.17\textheight]{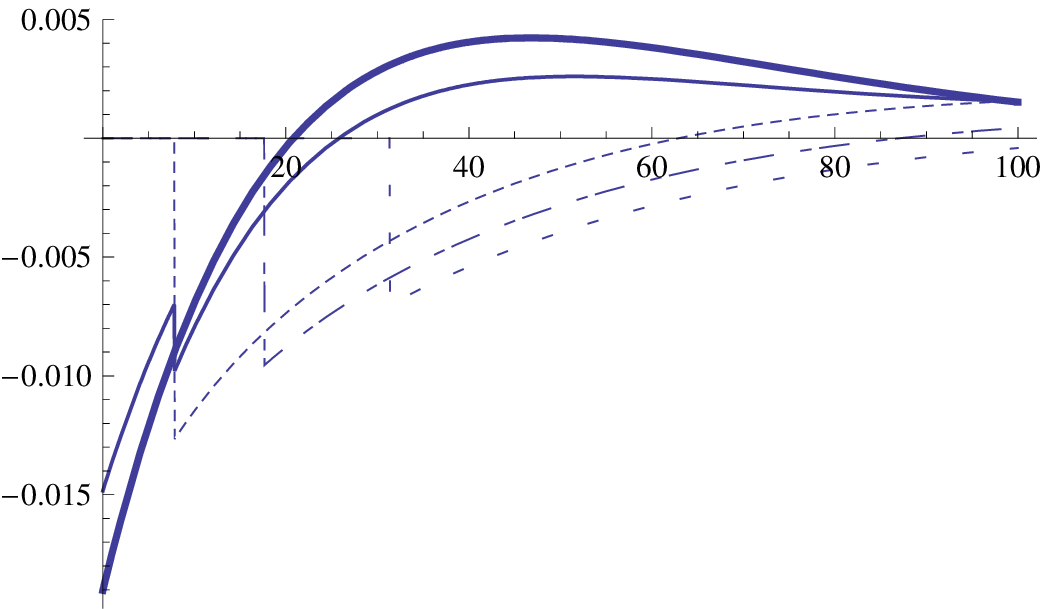}
}
\subfloat[]
{
\rotatebox{90}{\hspace{0.0cm} $d{\tilde H}/dQ\rightarrow$kg/(y keV)}
\includegraphics[height=.17\textheight]{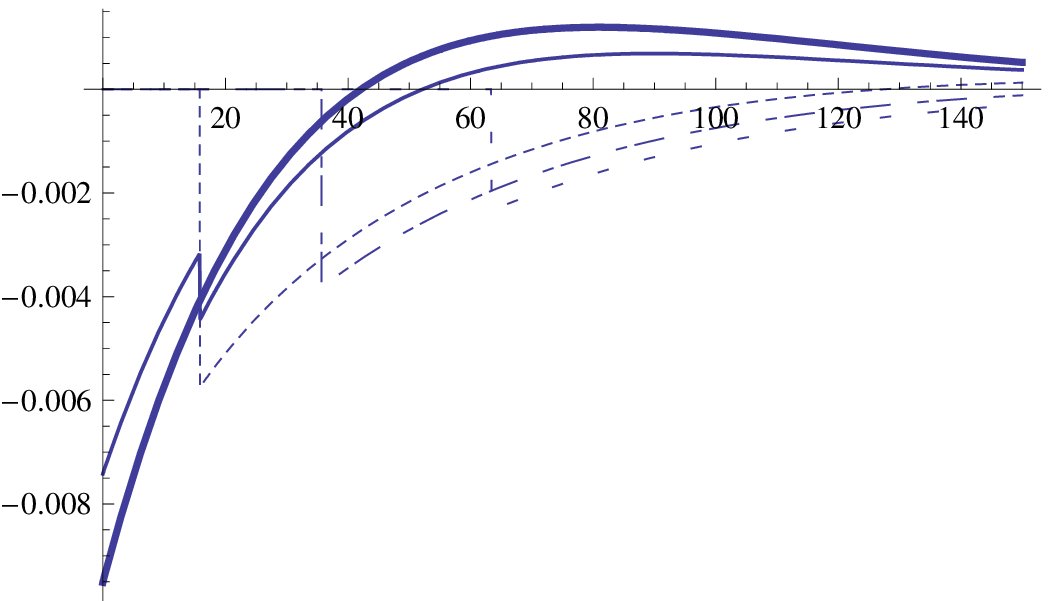}
}
\\
{\hspace{-2.0cm} $Q\rightarrow$keV}
\end{center}
\caption{ The differential rate $\frac{d\tilde{H}}{dQ}$,   as a function of the recoil energy for an intermediate target, e.g. $^{73}$Ge assuming a nucleon cross section of $10^{-7}$pb. Panels (a) (b), (c) and (d) correspond to to 5, 20, 50 and 100 GeV WIMP masses. Otherwise the notation is the same as that of Fig. \ref{fig:flowv}.
 \label{fig:dHdQ73}}
\end{figure}

 The introduction of debris flows  makes a small contribution at very low  energy transfers. As expected \cite{spergel12}, it tends to be favored compared to the M-B distribution as the recoil energy  increases. 
  This is particularly true for  small WIMP-nucleus reduced mass (see Figs \ref{fig:dRdQ127}, \ref{fig:dRdQ23} and \ref{fig:dRdQ73}). One does not see any particular signature in the shape of the resulting curve, simply all the differential rates fall with the energy transfer.
 One, however, observes an interesting pattern concerning the modulation amplitude, which affects time varying (modulated) part of the rate (see  Figs \ref{fig:dHdQ127}, \ref{fig:dHdQ23} and \ref{fig:dHdQ73}). We first focus on the small recoil energy region. Here we note that the sign of the modulation amplitude due to the flows is opposite to that of the M-B distribution for low reduced mass. For a light  target this occurs almost with any WIMP mass and extends up to energies of 100 keV.  Thus the use of a light target nucleus, like  $^{19}$F employed by  PICASSO \cite{PICASSO12} with the claimed achievement of 1.7 keV threshold, assuming that they will be able to detect the time variation of the rate, may differentiate between standard dark matter and flows. This can also be achieved by experiments involving heavier targets, like $^{127}$I or $^{131}$Xe, or $^{73}$Ge, assuming  a threshold of less than 1 keV, but, unfortunately, only  if the, yet unknown,  WIMP mass happens to be sufficiently small.
 
 The low mass target also exhibits interesting behavior at the high recoil energy  region. Indeed the modulation amplitude  due to the flows increases and becomes fairly large and positive, while that for the M-B distribution tends to decrease and becomes quite small. Again this behavior is pretty much independent of the WIMP mass. 
  On the other hand, for medium mass or heavy targets we expect a similar behavior  only in the case of light WIMPs.\\
In view of the above,  it seems to us that experiments on light targets, assuming that they overcome the fact that the expected coherent rates are smaller,  may provide more information than 
 simply challenging or confirming the claims
of seasonal modulations by the DAMA \cite{DAMA11} and CoGeNT \cite{CoGeNT11} experiments. We also expect, that the behavior of the relative modulation amplitude in the case of the spin contribution will be similar to that found here for the coherent rates. Then, the light odd targets are not disfavored from this point of view.

 Before concluding this section, we should mention that, even though we have exhibited in our figures the  modulation amplitude in absolute units (events per kg target per year), to get 
the time variation of the rate one should multiply this amplitude with $\cos{\alpha}$. As we have already mentioned,   the location of the maximum  depends on the sign of this amplitude.
 
 The above results, as we will see in the next section,  have important implications on the total event rates.

\section{Some results on total rates}

For completeness and comparison, we will briefly present our results on the total rates. Integrating the 
differential rates discussed in the previous section we obtain the total rate $R$ by adding the corresponding time averaged rate $R_0$ and the total modulated rate $\tilde{H}$, i.e.:
\beq
R=R_0+\tilde{H}=\frac{\rho_{\chi}}{m_{\chi}}\frac{m_t}{A m_p}  \left ( \frac{\mu_r}{\mu_p} \right )^2 \sqrt{<\upsilon^2>} A^2 \sigma_n t\left (1+h \cos{\alpha}\right ) ,
\label{Eq:Trates}
\eeq
with
\beq
t=\int_{Q_{th}/Q_0(A)}^{(y_{\mbox{\tiny max}}/a)^2}\frac{dt}{du}du,\quad h=\frac{1}{t}\int_{Q_{th}/Q_0(A)}^{(y_{\mbox{\tiny max}}/a)^2}\frac{dh}{du}du.
\label{Eq:thfac}
\eeq
 $ y_{\mbox{\tiny max}}$ is the maximum velocity allowed by the distribution ($y_{\mbox{\tiny max}}= y_{\mbox{\tiny esc}}$ in the case of the M-B distribution), and $Q_{th}(A)$ is the energy cut off imposed by the detector.
 
\begin{figure}
\begin{center}
\subfloat[]
{
\rotatebox{90}{\hspace{0.0cm} $R_0\rightarrow$kg/y}
\includegraphics[height=.17\textheight]{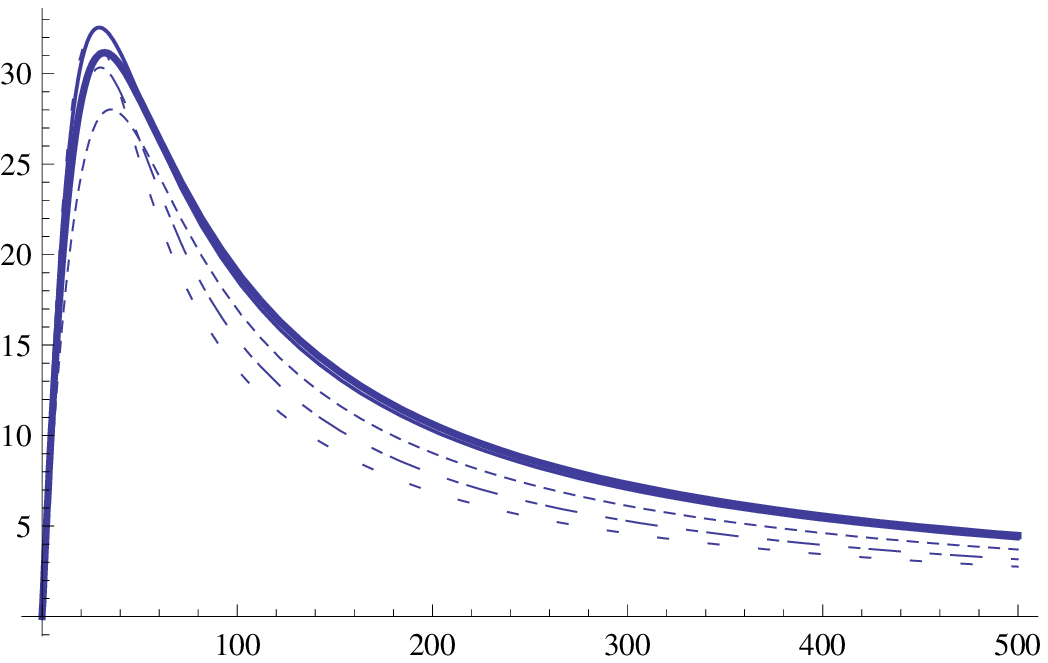}
}
\subfloat[]
{
\rotatebox{90}{\hspace{0.0cm} $R_0\rightarrow$kg/y}
\includegraphics[height=.17\textheight]{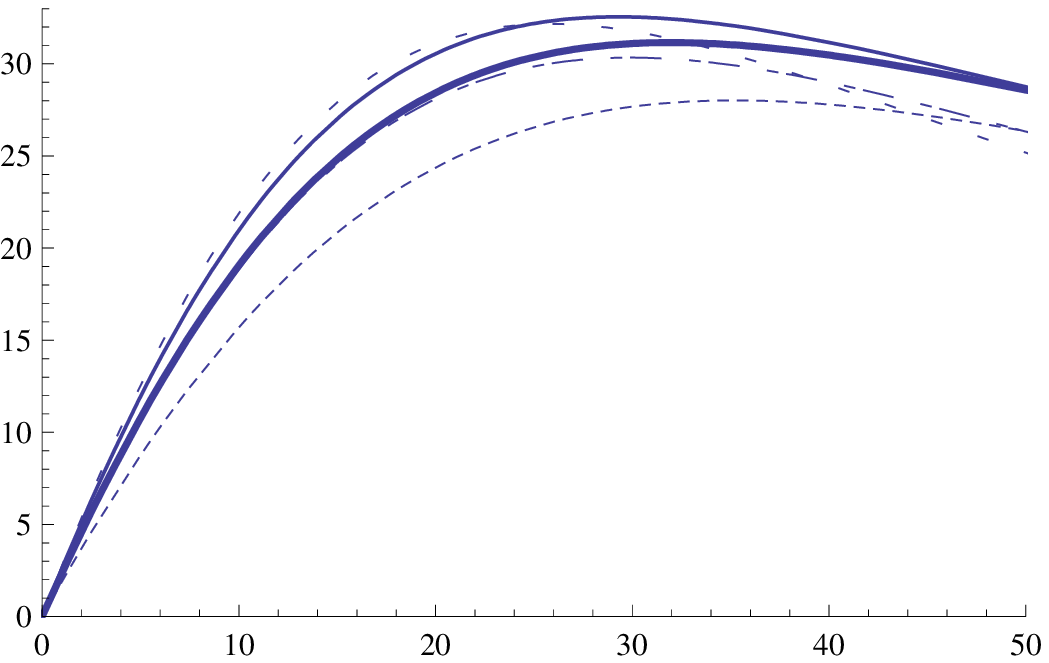}
}
\\
\subfloat[]
{
\rotatebox{90}{\hspace{0.0cm} $h\rightarrow$}
\includegraphics[height=.17\textheight]{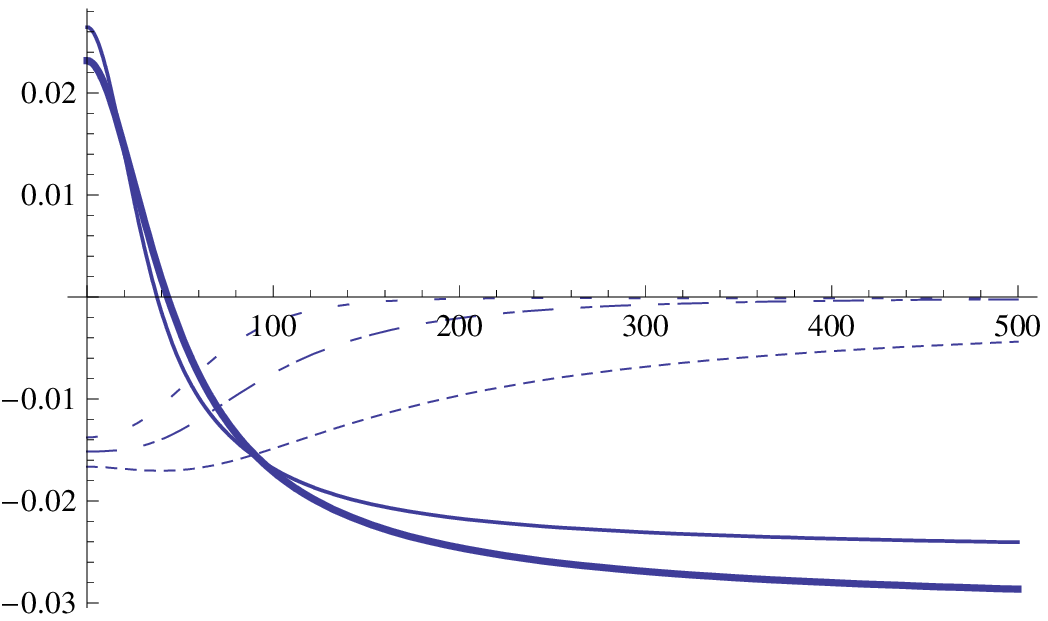}
}
\subfloat[]
{
\rotatebox{90}{\hspace{0.0cm} $h\rightarrow$}
\includegraphics[height=.17\textheight]{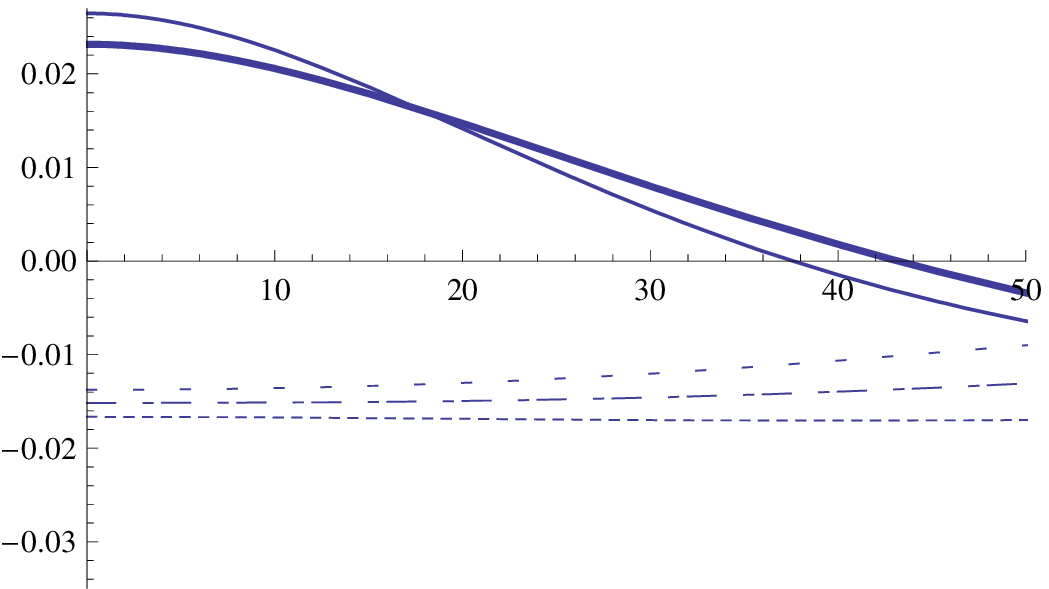}
}
\\
{\hspace{-2.0cm} $m_{\mbox{\tiny{\tiny  WIMP}}}\rightarrow$GeV}
\end{center}
\caption{ The total rate $R_0$ (top panels) and the relative modulation amplitude $h$ (bottom panels) as a functions of the WIMP mass in GeV in the case of a heavy target, like  $^{127}$I, at zero threshold. Note that the panels on the right column are a zoomed-in version of the corresponding ones on the left to better exhibit their behavior at the small WIMP mass region.  Otherwise, the notation is the same as that of Fig. \ref{fig:flowv}.
 \label{fig:Rh131}}
\end{figure}

\begin{figure}
\begin{center}
\subfloat[]
{
\rotatebox{90}{\hspace{0.0cm} $R_0\rightarrow$kg/y}
\includegraphics[height=.17\textheight]{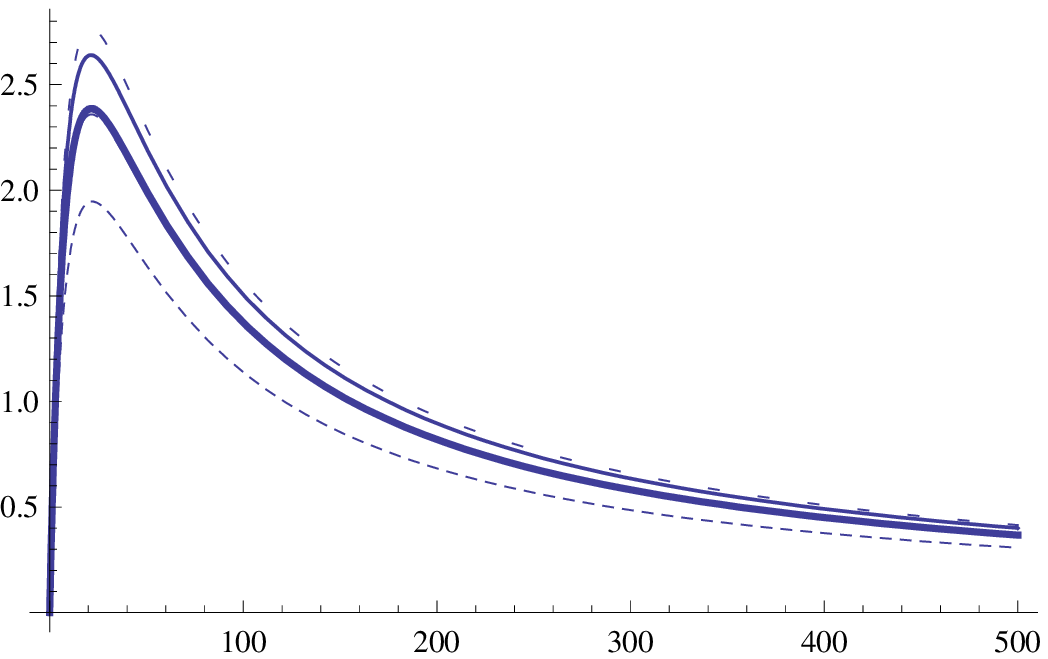}
}
\subfloat[]
{
\rotatebox{90}{\hspace{0.0cm} $R_0\rightarrow$kg/y}
\includegraphics[height=.17\textheight]{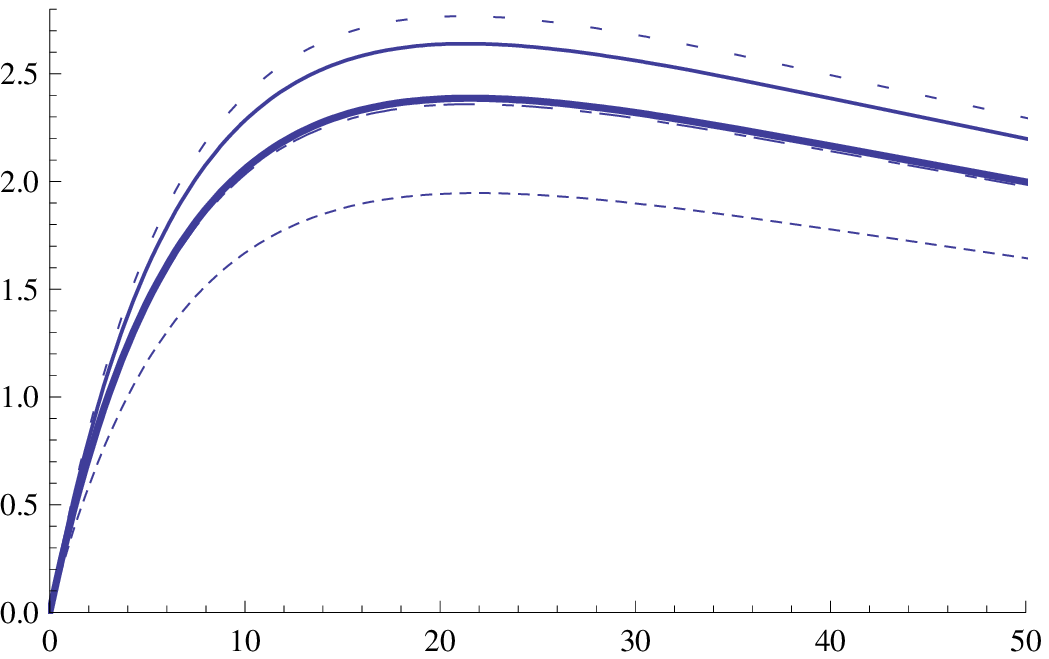}
}
\\
\subfloat[]
{
\rotatebox{90}{\hspace{0.0cm} $h\rightarrow$}
\includegraphics[height=.17\textheight]{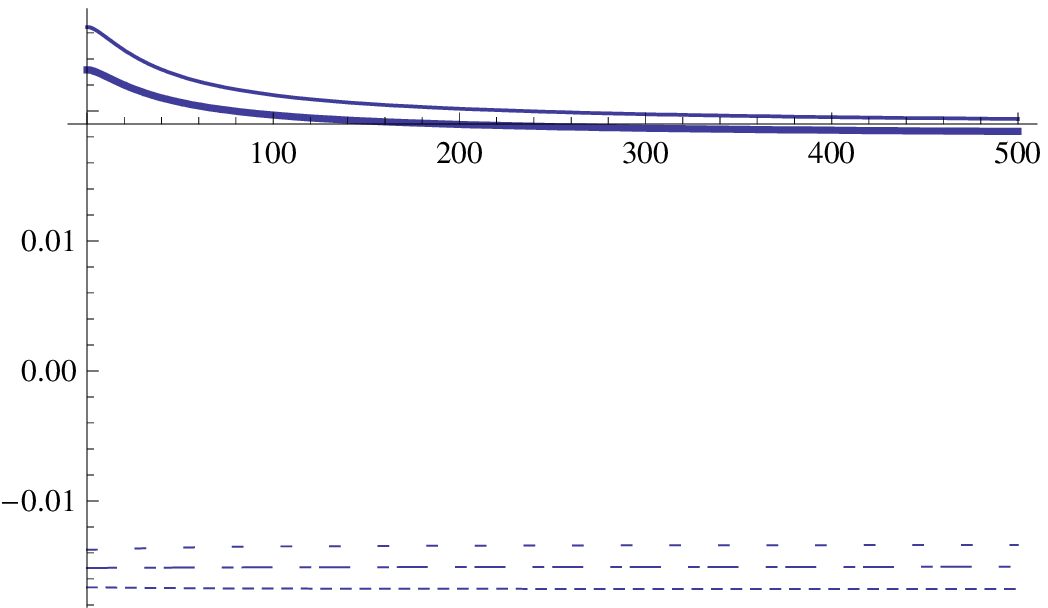}
}
\subfloat[]
{
\rotatebox{90}{\hspace{0.0cm} $h\rightarrow$}
\includegraphics[height=.17\textheight]{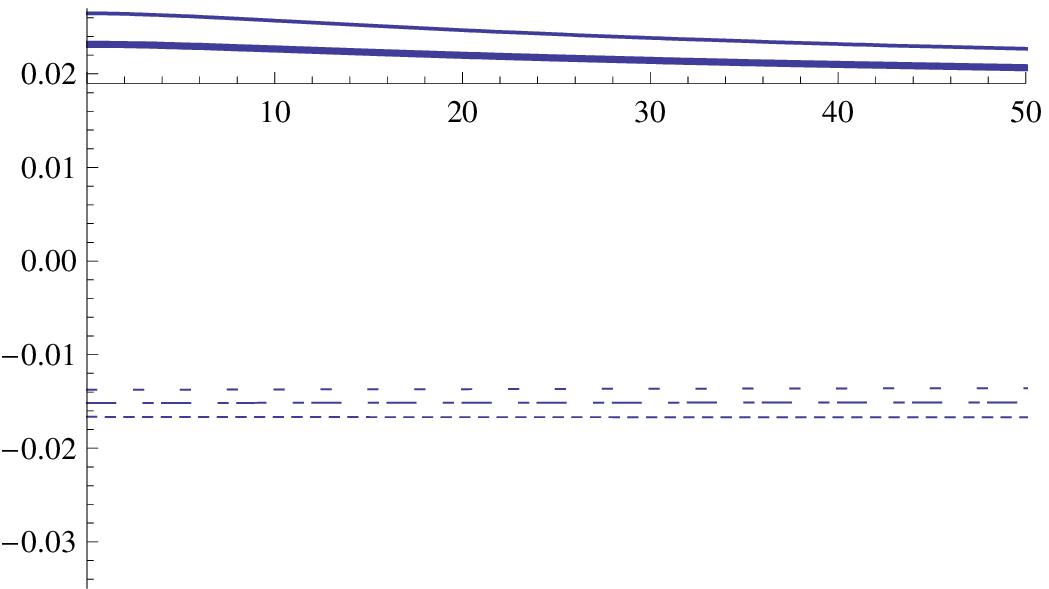}
}
\\
{\hspace{-2.0cm} $m_{\mbox{{\tiny WIMP}}}\rightarrow$GeV}
\end{center}
\caption{The same as in Fig. \ref{fig:Rh131} for a light target, e,g. $^{23}$Na. 
 \label{fig:Rh23}}
\end{figure}

\begin{figure}
\begin{center}
\subfloat[]
{
\rotatebox{90}{\hspace{0.0cm} $R_0\rightarrow$kg/y}
\includegraphics[height=.17\textheight]{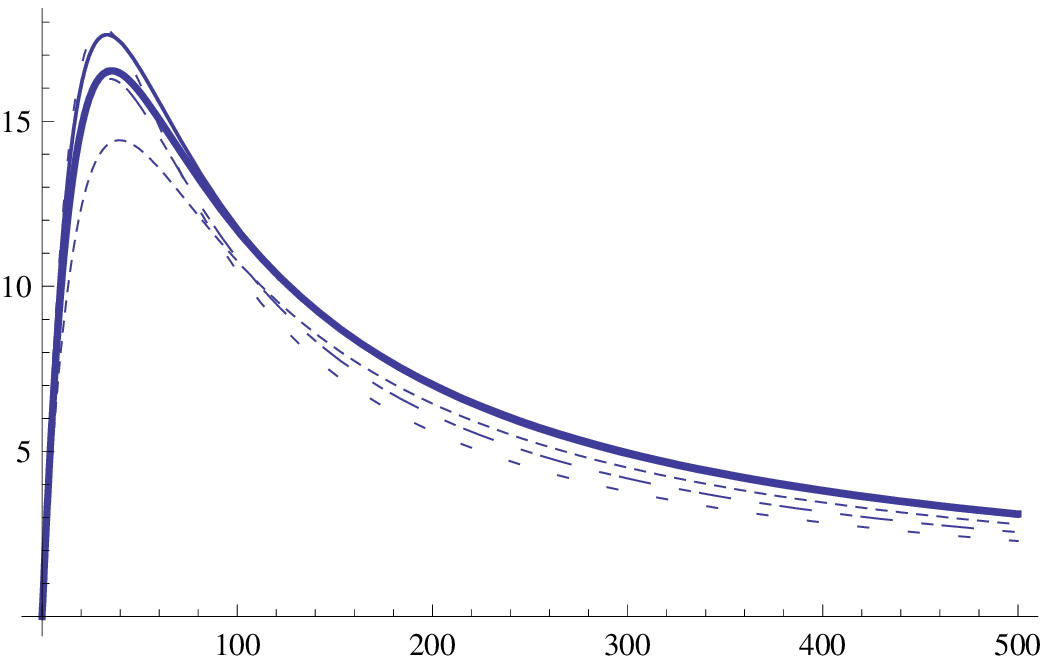}
}
\subfloat[]
{
\rotatebox{90}{\hspace{0.0cm} $R_0\rightarrow$kg/y}
\includegraphics[height=.17\textheight]{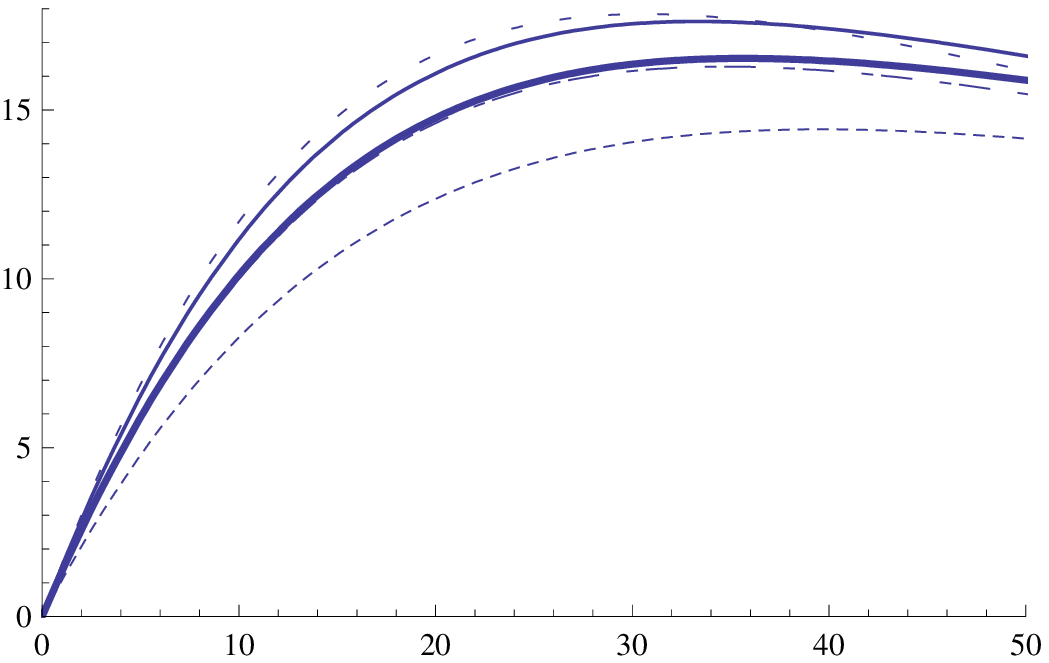}
}
\\
\subfloat[]
{
\rotatebox{90}{\hspace{0.0cm} $h\rightarrow$}
\includegraphics[height=.17\textheight]{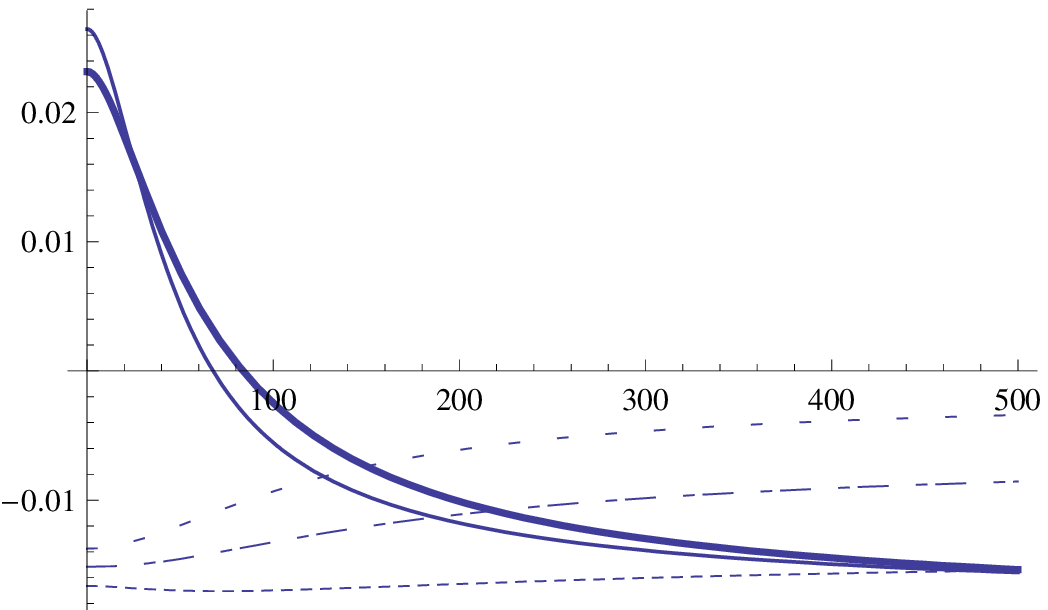}
}
\subfloat[]
{
\rotatebox{90}{\hspace{0.0cm} $h\rightarrow$}
\includegraphics[height=.17\textheight]{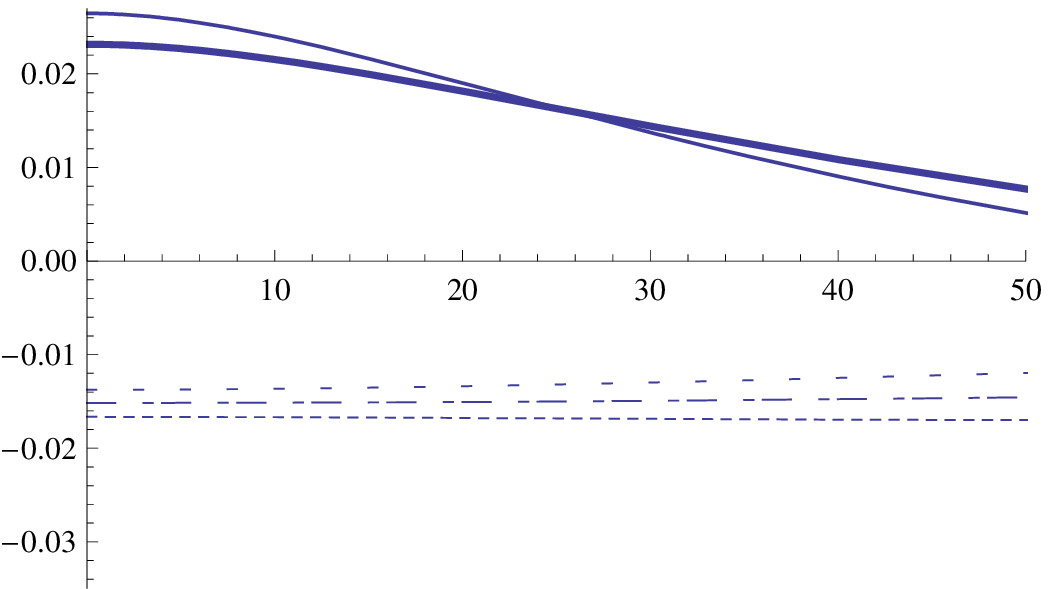}
}
\\
{\hspace{-2.0cm} $m_{\mbox{{\tiny WIMP}}}\rightarrow$GeV}
\end{center}
\caption{The same as in Fig. \ref{fig:Rh131} for an intermediate target, e,g. $^{73}$Ge.
 \label{fig:Rh73}}
\end{figure}

 The obtained results for the quantities $R_0$ and $h$ are exhibited in Figs \ref{fig:Rh131}-\ref{fig:Rh73} assuming a nucleon cross section of $10^{-7}$pb. In the case of a heavy target, the average event rate attains the maximum value of 30 events per kg of target per year at a WIMP mass of 30 GeV, while for heavy WIMPs it eventually falls to about 5 kg/y at 500 GeV.  For an medium weight target we get 15 kg/y at 30 GeV, with an asymptotic value of 4 kg/y. For a light target the maximum becomes 2.5 kg/y at 20 GeV. Again the asymptotic value at 500 GeV is about 1/5 of the possible maximum. This behavior of $R_0$ for WIMPs of large mass is easily understood by noting that the parameter $t$ essentially depends on the reduced mass. The rate $R_0$, however, contains an additional mass dependence, inversely proportional to the WIMP mass, arising in going  from the WIMP density  in our vicinity to the number density. Since for heavy WIMPs the reduced mass essentially becomes constant, equal to the target nuclear mass, the total rate, to a good approximation,  falls in this case inversely proportional to the WIMP mass. Similarly $h$, being the ratio of two quantities, depends only on the reduced mass and, thus, becomes essentially  constant in the high WIMP mass region.
   
  It is clear that, as far as the time averaged rates $R_0$ are concerned, the debris flows do not exhibit any characteristic signature to differentiate them from the standard M-B distribution. The relative modulation  amplitude $h$, however, exhibits a very interesting feature, namely, if caused by the flows themselves, it is negative for all targets, even for the light ones, and in the entire WIMP mass range (minimum in June).
 On the other hand, if it is caused by the M-B distribution, it is positive in the case of  light targets regardless of the  WIMP mass. It is also positive  for intermediate/heavy targets, if the WIMPs are relatively light. Then the maximum occurs on June 3nd as expected. It becomes negative only for relatively heavy WIMPs. This distinction is, however, washed out, if one  compares the case of the standard WIMPs on one hand with the combination of flows  and the  M-B distribution, in the form considered here,  on the other (compare the thick and the fine solid curves of panels (c) and (d) of Figs \ref{fig:Rh131}, \ref{fig:Rh23} and \ref{fig:Rh73}).

 
\section{Discussion}
In the present paper we first obtained results on the differential event rates, both modulated and time averaged, focusing our attention on the effects of debris flows. We found that:
\begin{itemize}
\item The flows indeed enhance the time averaged rates at relatively high energy transfers compared to the M-B distribution. All rates, however, fall as the energy transfer increases. This fall is only partial due to the velocity distribution. It is also caused by the nuclear form factor, especially in the case of heavy targets.
\item In view of the dependence of above rates on the unknown WIMP mass, from the time averaged rates one does not have a clear signature to  differentiate the debris flows from the standard distribution.
\item The differential modulated rates provide such a signature, the sign of the modulation amplitude, which determines the position of the maximum. The debris flows tend to favor a negative sign (minimum on June 3nd), while the standard WIMPs favor a positive sign when the target is light or even when the target is heavy but the WIMP is light (maximum on June 3nd). For small reduced masses such rates due to the debris flows tend to increase with the recoil energy and eventually they dominate over the M-B distribution. 
\end{itemize}
We then proceeded and calculated the total event rates as functions of the WIMP mass. We presented here results obtained with a zero threshold energy. For higher threshold energies we expect the debris flows to be suppressed a little less than the standard WIMPs \cite{JDV12n}, since the differential event rates associated with the former attain smaller values at low-energy transfers. The time averaged rates are affected by the debris flows, but one does not find a characteristic feature to differentiate the debris flows from the standard WIMPs.  The relative modulation  amplitude $h$, however, exhibits a very interesting feature, namely, if caused by the flows themselves, it is negative for all targets, even for the light ones, and in the entire WIMP mass range (minimum in June). On the other hand, if it is caused by the M-B distribution, it is positive in the case of  light targets regardless of the  WIMP mass. It is also positive  for intermediate/heavy targets, if the WIMPs are relatively light. Then, the maximum occurs on June 3nd as expected. It becomes negative only for relatively heavy WIMPs. This important distinction is, however,  washed out, if one  considers the  combination of flows  with the standard   M-B distribution in the manner considered here. Thus the   measurement of  the time dependence of the total event rate, with a relative difference between the maximum and the minimum of $2h\approx4\%$, may give a hint about the size of the WIMP mass. 

In conclusion, we have found that the measurement of the modulation, both in the differential and the total rates, for both light and heavy targets will shed light  i) on the mass of the WIMPs and ii) on the existence of flows. To this end, the differential event rate contains more information and may be a better discriminator. These issues may perhaps be settled even better, if data on  directional experiments become available.  Such theoretical explorations are currently under study. 
\section*{Acknowledgments} 
The author is indebted to the PICASSO collaboration for partial support of this work and to Viktor Zacek and Ubi Wichoski for their kind hospitality in Montreal and SNOWLAB and to Dr T.S. Kosmas for a careful reading of the manuscript and useful comments. 
This work was also partially supported by UNILHC PITN-GA-2009-237920.
\section*{References}

\end{document}